\documentstyle[  
               aps,psfig]{revtex}
\twocolumn
\narrowtext
\begin{document}

\draft

\title{Electrons in High-$T_c$ Compounds: Ab-Initio Correlation Results}

\author{Gernot Stollhoff}

\address{
Max-Planck-Institut f\"ur Festk\"orperforschung               \\
Heisenbergstra\ss e 1,  D-70569 Stuttgart, Germany            \\
}
\maketitle
\begin{abstract}
Electronic correlations in the ground state of an
idealized infinite-layer high-$T_c$ compound are computed using 
the ab-initio method of local ansatz. 
Comparisons are made with the local-density approximation (LDA)
results, and the correlation functions are analyzed in detail.
These correlation functions
are used to determine the effective atomic-interaction parameters for
model Hamiltonians.
On the resulting model, doping dependencies of the
relevant correlations are investigated. Aside from the expected
strong atomic correlations, particular 
spin correlations arise.
The dominating contribution is a strong nearest
neighbor correlation that is Stoner-enhanced due to the
closeness of the ground state to the magnetic phase. 
This feature depends moderately on doping, and
is absent in a single-band Hubbard model.
Our calculated  spin correlation function
 is in good qualitative agreement with that determined from
the neutron scattering experiments for  a metal.

\end{abstract}

\pacs{74.72.-h, 71.45.Gm, 71.15.-m, 75.10.Lp, 75.25.+z, 31.15.dv, 31.25.Eb}

\section{  Introduction    }             \label{sec:intro}
  
The theoretical understanding of the microscipic electronic properties
of the high-$T_c$ compounds
is still incomplete. The only ab-initio methods that so far 
have been applied to  these
compounds are based on the local-density approximation (LDA) within
the framework of
the density functional formalism \cite{LDA1,LDA2}. These 
fail to describe some of the
basic properties like the magnetic transition or the
magnetic correlations 
(for a review see \cite{pickett}).
Consequently, simplified models have been used that are mostly
restricted to a single band of strongly correlated electrons, and show
a Mott-Hubbard localization transition at half filling.
These seem to explain some of the magnetic properties but their
microscopic connection to the full Hamiltonian has not yet
been fully established (for a review see \cite{dagotto}). 

Here, we present the first application of the local ansatz (LA) to these
materials.
The LA is an
ab-initio method for the treatment of the correlated electronic
ground states of solids \cite{stol3,stoln}. 
It contains no
homogeneous-electron-gas like approximation wherever, 
and consequently has no problems in overlooking magnetism.
The LA yields not only ground-state energies or densities but
also detailed correlation functions. 
In particular we present the detailed intraplanar correlation features 
relevant for all high-$T_c$ compounds.
Of specific interest are the magnetic correlations.
We compare the frequency integrated momentum dependent
inelastic magnetic neutron scattering intensity measured for
$La_{0.85}Sr_{0.15}Cu_2O_4$ \cite{hayden} to the LA
results and connected it to a specific correlation.
\\

The LA 
is similar to  Quantum Chemistry (QC) methods,
which provide a satisfying description of electrons in small molecules.
It allows to extend the QC-accuracy to solid calculations.
Like most of these methods, the LA
adds correlations as corrections to a 
single-particle self-consistent-field (SCF) ground state
obtained from a Hartree-Fock  (HF) calculation where the electrons
are described in a restricted single particle basis
of Gaussian type orbitals (GTO's). 
The HF-computation for the solid is performed by the program Crystal92
\cite{crys}. At present,
the LA is the only ab-initio correlation scheme available
that makes use of this HF program..
Unlike the QC-methods,
the LA
allows a quantitative
treatment of electronic correlations for solids,
independent of their nature, i.e. whether they are
insulators or metals.
The LA can do so because
it
does not attempt to cover 
the complete spaces of one- or two-particle excitation
operators  as QC-methods usually do,
but considers the local character of the relevant correlations
from the very outset. It can be seen as an apropriate generalization
of the Jastrow-ansatz \cite{jast} to inhomogeneous systems.

Every correlation operator  in the LA scheme has a very specific meaning.
It is constructed from pairs of
local orbitals, each of which is connected with a single atom.
Consequently, all the incorporated 
correlation corrections are separated into those
on single atoms and those inbetween atom pairs.
The full correlation treatment can be segmented and partitioned
with the individual atoms as the smallest available subunit. The
correlation operators and their treatment are essentially independent of the 
nature of the SCF-ground state, i.e. whether this be metallic or otherwise.
Such a restrictive choice for correlation operators leads to a
strong reduction in the correlation-operator space and thus
substantially facilitates
computations.
Necessarily, it admits 
a small loss of the correlation energy available in a full treatment 
within a given basis set. From previous calculations, this loss is known
to be only 2\%, independent of the system size \cite{stoln}.

The LA was used before for extended molecules like
$C_{60}$\cite{sh}, three dimensional semiconductors \cite{diam}
and ionic insulators \cite{stol6}, as well
as one-dimensional (polyacetylene \cite{koe2}), two-dimensional (graphite
\cite{grap}),
and three-dimensional metals (Li\cite{andrea}). The calculations
presented in this work concern the first application of the LA
to a metallic transition-metal compound.
\\

From the ab-initio results, also 
model Hamiltonians can be derived 
that are based on atomic
degrees of freedom. The ab-initio
correlation functions obtained from the LA
allow in particular to unequivocally determine
the model
interaction parameters \cite{8}. This feature enables us to extend correlation
calculations to problems that are still out of reach for ab-initio 
LA calculations, and further to make comparisons to models determined
by other methods. This is particular relevant 
to the high-$T_c$ compounds, whose properties are often adressed by
means of model calculations.

Models based on atomic degrees of freedom consist of
selected sets of orthogonalized atomic orbitals $i$ that are related
by hopping terms. The interaction part
of the Hamiltonian, $H_{int}^{mod}$, is usually restricted to  local
interactions 
\begin{equation}  
H_{int}^{mod}=\sum_{i} U_i n_{i \uparrow}n_{i \downarrow}
\end{equation}
for two electrons in the same orbital $i$. Each 
interaction-energy parameter $U_i$ dominantly
influences a particular correlation function of the correlated
ground state $\Psi_{corr}$ for such a model, namely
$\langle \Psi_{corr}| n_{i \uparrow}n_{i \downarrow}|\Psi_{corr}\rangle$, or
equivalently the change of this correlation function due to correlations,
\begin{equation}  
\Delta_i(corr)=\langle \Psi_{corr} |n_{i \uparrow}n_{i \downarrow} 
|\Psi_{corr}\rangle-
\langle \Psi_{SCF} |n_{i \uparrow}n_{i \downarrow}|\Psi_{SCF}\rangle
\ \ \ .
\label{fluct}
\end{equation}
Here,  $\Psi_{SCF}$ represents the SCF-ground state of the model.
As for the ab-initio treatment, the model ground state is also
computed by means of the LA.
$\Delta_i(corr)$ equals zero for $U_i=0$ and rises continuously 
with $U_i$.
When multiplied with $U_i$, it is a measure of
the interaction energy. This is the relevant relative quantity,
when symmetry is broken due to the atomic interactions, and a
magnetic or structural phase transition occurs.
 In such a case, states are compared
that differ in those atomic fluctuations.
These  energy costs due to local charge fluctuations are also
relevant in the context of Compton scattering,
secondary or shake up peaks in photo emission
or core spectroscopy. For the transition metals,
it turns out that the same model interactions
are needed for the description of all these properties 
\cite{tra2,soh,saw,sco}.

With the unequivocally defined orthogonalized atomic orbitals available
in the ab-initio calculation (see section \ref{sec:scheme}),
this same correlation
function can  be determined from the ab-initio calculation.
The model interaction $U_i$
can thus be fixed by demanding that the corresponding
model correlation corrections  $\Delta_i(corr)$ agree to the same ab-initio 
quantities. 
This connection can also be used to analyse screening details entering
such a
model interaction. By adding stepwise  particular
screening contributions in the ab-initio calculations,
we will determine how the model interaction arises,
 starting from the
bare Coulomb interaction. 
Such an investigation was in the past performed
for $\pi$-electron interactions in organic
compounds \cite{8}.

For the high-$T_c$ materials, ab-initio correlation functions have not
yet been available. Model interactions, however have been computed from
the LDA calculations in a different approximation, by freezing
specific charges on individual atoms and relaxing the environment. 
For the most extended model that has been
used in this context, a three band model,
the two interaction parameters, namely an effective local interaction $U_d$
between the $Cu 3d$-electrons, and an interaction $U_p$ between the 
$O 2p$-electrons have been obtained in this way \cite{mms,sj,hyb}.
These interactions have 
then been used to extend the LDA calculations to magnetic properties
(by means of the so-called
LDA+U method or by the self interaction correction 
(SIC) calculations) \cite{zaa}. 
Model calculations based on these interactions yield photoemission
results that are 
in very
good agreement with experiment \cite{gunnpe}.
This agreement is in contrast to the case of in particular the middle of
the transition metal series
where effective interactions obtained by the LDA do not match 
the interactions needed for the description of the above-mentioned
properties of these systems (for a comparison, see \cite{8}). 
For the high-$T_c$ compound that we shall deal with, we will make
a comparison between the effective interactions
obtained by the LA and by the LDA. It is the first of its kind
because for the systems treated so far
by the LA no LDA interactions are available. The comparison will shed 
additional light on the transition-metal case.

For simpler single-band models, interactions
have  so far been mostly guessed, or obtained by means of
fits to specific experiments.
Often, they are deduced from the LDA three-band models. However,
usually the strong interaction found by the LDA is the only transferred
quantity. The $Cu 3d_{x^2-y^2}$-occupation of 1.5 of the LDA \cite{zaa}
was usually replaced by occupations smaller than 1.2 \cite{dagotto}.
Only under such conditions, a Mott-Hubbard scenario applies \cite{host}.
\\

This paper is organized as follows. In section \ref{sec:scheme}, 
a short description of
the LA and of its possible shortcomings in connection with its
application to high-$T_c$ compounds is given. Section \ref{sec:imp}
contains a detailed discussion  of the ab-initio computation 
and an analysis of the correlation functions.
These calculations are performed for
an idealized high-$T_c$ compound, the so called infinite-layer system.
Whereever possible, comparison to the LDA results is made.
Sections \ref{sec:mod} and \ref{sec:mod3}
 deal with the computation of the
model interactions. The former section contains in particular a 
detailed account of the screening
mechanisms for the $3d$ transition-metal interaction.
In section \ref{sec:mod4}, the  
dependencies of the most important
correlation features on band filling are investigated.
This analysis is made on the model level.  The
ab-initio program Crystal92 that is used for the SCF calculation is
restricted to integer electron occupations per unit cell, and a 
small change of band filling
on the ab-initio level would demand large unit cells. 
Finally in section \ref{sec:mod5}, a comparison 
to neutron-scattering results for a metallic
compound is made, and the experimentally found inelastic scattering is 
connected to a particular correlation obtained by the LA calculation. 
A summary of our work is presented in the
concluding section \ref{sec:conclu}.

     \section{Computational details}     \label{sec:scheme}

The aim of the paper is a quantitative understanding of the electronic
correlations relevant to all high-$T_c$ compounds, namely those
in the individual $CuO$-plane. To simplify the computations, the idealized 
compound $SrCuO_2$ is treated. It is a so called infinite
layer compound with $CuO$-planes separated by layers of $Sr$ ions. In the
planes, the
$Cu$-atoms form a quadratic net with a separation of 3.925 \AA, with the 
$O$-atoms
at equal distances inbetween neighbor $Cu$-atoms.
The stacking distance, and the perpendicular $Cu$-separation
amounts to 3.43 \AA. The $Sr$-ions have equal distance to four $Cu$-atoms in 
the two neighbor planes each. 
This compound has the smallest possible unit cell 
containing four atoms. The
uppermost valence band is half filled. Therefore, the compound is
expected to order
antiferromagnetically. Our interest , however, is in the 
correlations in the metallic state. To represent 
a doped metallic compound would require a very much larger unit cell. 
Instead, the metastable non magnetic state for the small unit cell
is used as a starting point.
This approximation is justified because
it is known from LDA calculations that the 
energy bands and Fermi surface
of the metastable half-filled metallic state are very close to the ones 
for the true metallic compounds \cite{pickett,aba}. 

In a first approximation, this metastable metallic state is
obtained from a restricted, non symmetry broken
Hartree-Fock calculation for this compound, performed with the ab-initio
program
Crystal92 \cite{crys}. 
For the $Cu$ and the $O$-atoms, good all-electron-GTO-basis sets are used. 
For $Cu$, this is a modified (14,11,6) Ahlrichs basis \cite{wae}, 
contracted to (6,4,2) orbitals. From the original basis, the outermost
diffuse functions were removed, and the next exponents adjusted 
and reoptimized.  For $O$. this is a (11,7) Huzinaga
 basis \cite{rs},  whose outermost exponents
were contracted as was done before \cite{diam,stol6}, plus a 
set of d-orbitals.
While the basis sets for the atoms in the planes are of good quality and
promise results for the valence electrons close to the Hartree Fock limit,
this is not the case for the $Sr$ atoms. 
The latter are represented by a large core pseudopotential and a 
single 5s orbital
\cite{hw}. Here, also the outermost diffuse basis orbital was removed. 
There is no need for such a treatment, but
due to this choice, the charge distribution
and correlation analysis can be definitely restricted
to the degrees of freedom within the plane.

From the SCF calculation, the metallic
single-particle ground state  $|\Psi_{\rm SCF}\rangle$ is obtained.
Its Fermi surface is identical to the one of a LDA-calculation
for the same compound \cite{aba}. Also the uppermost energy band is
similar to the equilavent LDA band, except of an additional homogeneous 
spreading by almost a factor of two due to the non-local and non-screened
exchange. The non-local exchange also causes the lower lying bands
to be more separated from the uppermost half-filled
 band than in the LDA calculation. 
A presentation of such HF bands, obtained in a
somewhat different basis, is found in Ref. \cite{mei}.

In a next step, correlations are added by the LA.
Here, the following variational ansatz is made
for the correlated ground state:

\begin{eqnarray}
| \Psi_{\rm corr} \rangle & = & e^{-S} | \Psi_{\rm SCF} \rangle 
\label{eq:lanst0} \\
      S                   & = &  \sum_{\nu} \eta_{\nu} O_{\nu}   
\label{eq:lanst01} \\
O_{\nu}                   & = &  \left\{ \matrix{
                                      n_{i \uparrow} n_{i \downarrow} \cr
                                      n_i n_j                         \cr
                                      \vec{s}_i \cdot \vec{s}_j     \cr
    \{n_{i \uparrow} (a_{i \downarrow}^{\dagger}a_{j\downarrow}-
a_{j \downarrow}^{\dagger}a_{i\downarrow})\} +
\{ \uparrow \leftrightarrow \downarrow \} \cr
n_i } \label{eq:lanst}
                                                     \right. \ \ .
\end{eqnarray}
The $\eta$'s serve as variational
 parameters. The $n_{i \sigma}$ and $\vec{s}_i$ are density and spin operators for an
electron in the local state $a_{i \uparrow}^{\dagger}$, represented by
the orbital
\begin{equation}
g_i(\vec{r}) = \sum_j \gamma_{i j} f_j(\vec{r})       \label{eq:local}
\end{equation}
where the $f_i(\vec{r})$ are the (GTO like) basis orbitals. 
The operators have an
obvious meaning. The first operator $n_{i \uparrow} n_{i \downarrow}$,
for example, when applied to $ | \Psi_{\rm SCF} \rangle $, projects out all
configurations with two electrons in orbital $ g_i(\vec{r})$. In connection
with the variational parameter $ \eta_{\nu}$, as in eq.~\ref{eq:lanst01}, it
partially suppresses those configurations. Similarly, the operators $n_i n_j$
describe density correlations between electrons in local orbitals
$g_i(\vec{r})$ and $ g_j(\vec{r})$. For the homogeneous electron gas, an ansatz
with these two kinds of operators leads to the Jastrow
function~\cite{jast}. The operators $ \vec{s}_i \cdot \vec{s}_j $ generate
spin correlations. The fourth kind of operators is of the form of
$[O_{\nu},H_0]_-$, where $H_0$ represents the single-particle Hamiltonian.
In comparison to the first three kinds of operators which look like
particular interaction contributions, these operators refine the
ansatz with respect to the band energy of the electrons involved \cite{stoln}. 
Within the computation, the original operators of eq. 
\ref{eq:lanst} are modified by subtracting the
 contracted contributions in each of them.
The corrected operators when applied to $ | \Psi_{\rm SCF} \rangle$
contain only two-particle excitations, and the corrected
last kind of operators in eq. \ref{eq:lanst} covers local single 
particle excitations, i.e. it allows for changes in occupations.

   The variational parameters $\eta_{\nu}$ are chosen to optimize
the energy
\begin{eqnarray}
E_G & = & \frac{\langle\Psi_{\rm corr} | H | \Psi_{\rm corr} \rangle}
               {\langle\Psi_{\rm corr} | \Psi_{\rm corr} \rangle} \\
    & = & \langle\Psi_{\rm corr} | H | \Psi_{\rm corr} \rangle_{c} \ \ .
\end{eqnarray}
In the last equation,
 the subscript ${}_{c}$ indicates that only connected
diagram contributions are summed up.
This expression cannot be evaluated exactly.
The standard approximation is an expansion in powers of $\eta$, up to second
order,
\begin{eqnarray}
E_G      & = & 
       E_{\rm SCF} + E_{\rm corr}                 \\
E_{\rm corr} & = &  - \sum_{\nu} \eta_{\nu} \langle 
O^{\dagger}_{\nu} H\rangle \\
     0 & =  &  - \sum_{\nu} \eta_{\nu} \langle O^{\dagger}_{\nu} H\rangle
     + \sum_{\nu,\mu} \eta_{\nu} \eta_{\mu} \langle O^{\dagger}_{\nu}
HO_{\mu}\rangle_c
     \ \ .               \label{eq:expan}
\end{eqnarray}
Here, $\langle A \rangle$ means the expectation value of the operator
$A$ within
$| \Psi_{\rm SCF}\rangle$. 
This approximation works
only if the correlations are sufficiently weak. 
Disregarding the reduced subspace of
correlation operators, the approximation used so far
corresponds to a Linearized Coupled Cluster Singlet and Doublet (LCCSD)
treatment \cite{cizek}. It can be
extended to a CCSD treatment \cite{cc}.

The local orbitals in eq.~\ref{eq:local}
are connected to a single atom only and are built from
its basis orbitals. This is the essential approximation of the LA. 

In the present application, only
atomic orbitals are constructed.
These are uniquely determined from the SCF-ground state
by the condition that
they are built from basis orbitals on the respective atoms only and
that they contain a maximal part of the occupied space.
The resulting orbitals are next L\"owdin-orthogonalized to each other.
More  localized subatomic orbitals that are usually included in applications
of the LA are not used in this first application
for a metallic high-$T_c$ compound.
Therefore, only interatomic correlations are treated, i.e.
correlations that are described by operators built from the atomic orbitals.
Shorter range or intra-atomic correlations, as well as particular
polarization correlations are not covered.
From previous experience, it is known that such contributions are not 
very relevant
for the topics of interest here.
Estimated corrections due to the omitted correlations will be given were
they are nonnegligible.
\\

In the LA, the correlations are taken into account incrementally.
The correlation energy is exactly expressed as an incremental sum
over contributions from different sets of atom clusters
\begin{eqnarray}
 \label{eq:corrtot}
E_{\rm corr}  & = & \sum_{m=1}^N \
                    \frac{1}{m!}
\left\{ \  \sum_{j_1}^N \sum_{j_2}^N \ldots \sum_{j_m}^N
\left\langle E_{\rm corr} \left(
A_{j_1} A_{j_2} \ldots A_{j_m} \right) \right\rangle_i
\right\}  \nonumber \\
          &  & \ \ \ \ \  {\rm with}\ \ j_1 \neq  j_2 \neq \ldots \neq j_m
\end{eqnarray}
where the $A_{j_n}$\ denote atoms, on and between which correlation
operators are formed and 
it holds , for example,
\begin{equation}
\left\langle E_{\rm corr} \left(A_1 A_{2}  \right)
\right\rangle_i\ =  E_{\rm corr}(A_1 A_2)
- E_{\rm corr}(A_1) - E_{\rm corr}(A_2)
\end{equation}
i.e.\ the 
increments $\langle \rangle_i$ include only the 
changes of the correlation energy due to the extended set represented.
Translation invariance and the particular local symmetry are easily
included by performing the above summation only over the subset of
symmetry inequivalent clusters.
For every computation,
the exact solid single-particle expectation values 
are taken. 

For the coverage of the interaction part,
the local nature of
the correlation operators allows a drastically simplifying reduction.
In a finite basis set per atom representation, the interaction is
represented by a fourth order tensor of basis interaction matrix elements
whose indices extend over
the involved basis orbitals. The generation of this tensor and its
handling are the limiting steps in a correlation calculation.
For the particular 
correlations on the set of atoms treated, the required tensor
can,
to a very good approximation, be restricted to these atoms plus all their
nearest neighbours. This restriction makes all
required computations easily feasible. Possible corrections due to lacking
matrix elements  are included in computations extending over larger clusters.
The two largest clusters for which explicit ab-initio
correlation calculations
were performed are depicted in Fig. \protect\ref{fig1}. The larger one consists
of five active $Cu$ atoms and four active $O$ atoms. The basis interaction 
matrix elements
are computed for the whole $Cu_5 O_{16}$ cluster. As will be
demonstrated, the most relevant informations can be satisfactorily
obtained from correlation calculations extending up to this size of clusters.
\\

Two approximations made for the handling of the LA 
need further discussion 
in connection with
the application to a high-$T_c$ compound. 
The one is the restriction to weak correlations
while the high-$T_c$ compounds are usually connected with a Mott-Hubbard   
transition.
The LCCSD approximation, in which the LA is computed, fails for the
strongly correlated half-filled band case. However it does so in a
controllable way. For too strong correlations, the
correlation
corrections turn too large and lead to negative density correlations
functions. The criterion of positive density correlations
can be taken as an indication whether the LA results are still meaningful.
Away from half-filling, the LA behaves better, and for
an almost empty band, it applies even for diverging interaction.
Also, a more quantitative test can be made. 
Individual correlation corrections like the one
due to a single correlation operator $n_{i\uparrow}n_{i\downarrow}$
can be computed variationally, i.e. exactly. Here, a comparison with the
result of the eqs. \ref{eq:expan} in the same operator subspace
can be made, and the overestimation
of the correlation expansion can be quantified.
Such
a variational calculation restricting to two-particle excitations
has no meaning for the treatment of the full extended system (N) due
to lack of size consistency.  
The resulting correlation 
energy would scale like $\sqrt{N}$.
When such
a variational computation is extended to  more then a single operator,
it can  only give a lower limit to the 
correlation results.
Earlier ab-initio
calculations with the LA for finite $Cu-O$ clusters \cite{mest} and
experience with succesful LA model calculations for the transition
metals themselves \cite{st,tra2,soh} have already indicated that the 
range of applicability of the LA might well extend to
the high-$T_c$ materials.

The second approximation made is the restricted SCF-ground state that is
used as a starting
point, although the 
particular system chosen is known to be antiferromagnetically ordered.
The LCCSD approximation is sensitive
to antiferromagnetic order.
This is in contrast to standard perturbation
expansions, and
can be seen by
resolving the LCCSD equations (\ref{eq:expan}), leading to
\begin{equation}
E_{\rm corr} =  - \sum_{\nu\mu} \langle 
O^{\dagger}_{\nu} H\rangle ( \langle O^{\dagger}
HO\rangle_c)_{\nu\mu}^{-1}\langle O^{\dagger}_{\mu} H\rangle
     \ \ .               \label{eq:enex}
\end{equation}
The denominator contains the exact two-particle excitation energies.
If the restricted SCF-ground state turns unstable, its susceptibility
diverges for a particular wave vector, and consequently 
two-particle excited states must exist with 
energies degenerate to or even lower than the
SCF ground state energy.
With sufficiently extended sets of
correlation operators, the matrix $\langle OHO\rangle$
is no more positive definite, and the scheme
turns instable.
From such a calculation, also the smallest set of correlation
operators may be determined that leads to instability.
In particular, a lower limit for the size of stable magnetically ordered
domains can be obtained. 
This holds true as long as the phase transition from
the metal to the insulator as a function of occupation
is second order. If the transition is  first order, then
the computation in the metastable metallic state is still 
relevant for the doped metallic state but might lack informations
about the  magnetic order.

 \section{Results of the correlation calculations}     \label{sec:imp}

We will next
present the ab-initio  results of the LA 
in separate  chapters for 
correlation energies, charge distributions and particular 
correlation functions. In all cases, partial correlations
are consecutively added, starting from atomic terms, and extending up to the
longest-range correlations included, namely the ones
between third-nearest-neighbor
$Cu$-atoms.

     \subsection{Detailed correlation energies}     \label{sec:en}

In Table \ref{tab31}, the contributions of the different
classes of operators
to the correlation energy are  displayed. 
The sets of operators are grouped into those on individual atoms
and those inbetween different atoms. For the latter cases, also the
fraction of spin operator contributions is given.
As expected, the largest
overall energy gain is due to the on site or atomic correlations. Here,
the largest part is from the
$Cu-3d_{x^2-y^2}$ operators. However, almost twenty
percent of the on site contributions is connected with a charge transfer that
will be discussed in more detail later.

There are two different kind of relevant longer range contributions.
One arises from correlation operators between neighbor $Cu$ and $O$ atoms.
Here, no specific contribution is dominating. Rather the very local atomic 
correlation hole generated by the atomic operators
is smoothly extended, adding ten
percent of the on-site correlation energies.
The second kind is connected with spin correlations between different 
$Cu$-atoms, and in particular with those inbetween electrons in the 
$Cu 3d_{x^2-y^2}$-orbitals. These dominate the nearest neighbor $Cu-Cu$
correlations and are exclusively responsible for the longer range terms.
The neighbor $Cu-Cu$ correlations will be later elaborated in more detail.
The longer range correlations are connected with the eventual formation of
long range magnetic order. Next nearest $Cu-Cu$ contributions to the energy
are as large
as nearest neighbor contributions, indicating that here problems with the
metastability of the nonmagnetic SCF-ground state begin to show up. 

While all
shorter range correlations were fully or almost completely converged with
respect to the series of clusters selected, this does not hold true anymore
for the
third nearest neighbor 
$Cu-Cu$-contributions. However, divergence doesn't yet appear.
This indicates that stable
antiferromagnetic correlations at half-filling need to coherently
extend over domains larger
than the clusters selected for the correlation computation.
 
When added, all these correlations represent an energy gain of roughly 
5$eV$ per unit cell.
All these correlations are due to binding, and the resulting correlation 
energy amounts to a
large fraction of the total binding energy.

     \subsection{Partial charge distributions}     \label{sec:di}

The partial charge distributions 
$n_i(\Psi)=\langle\Psi| \sum_{\sigma}n_{i\sigma}|\Psi\rangle$
are presented in table \ref{tab32} for different states $\Psi$.
The first row contains the values for $\Psi=\Psi_{SCF}$.
When added, the partial occupations reach the
number of valence electrons up to 0.02. This indicates
the good quality of the computed
orthogonal atomic orbitals.
The occupation of the
$Cu 3d$-orbitals is very close to the estimate for the solid, obtained from
earlier finite cluster HF-calculations \cite{mest}, 
except for
the $Cu 4s,4p$-occupations.
In the earlier calculation, these came out smaller for two reasons.
The one is that a basis set was used that lacked
the most
extended exponents used here for the $4s,p$-orbitals.
This restriction was made
to avoid artifacts, resulting
from the large negative charging of the small clusters treated.
The second reason is that
in the earlier calculation,
the $4s$ and $4p$-orbitals were Schmidt 
orthogonalized to the $O 2s,2p$-orbitals, while here
all orbitals are equally treated by a mutual L\"owdin orthogonalization. 
\\

With the addition of correlations, a relatively large charge transfer 
occurs. Ultimately, it is a charge transfer mostly from the
$Cu 3d_{x^2-y^2}$-orbitals into the $O 2p$-orbitals. However, for
its understanding it is necessary to progress stepwise.
A first step is the addition of atomic correlations which
lead to a large correlation energy gain.
The dominant charge transfer due to the atomic correlations
is from the $Cu 3d_{x^2-y^2}$-orbitals
to the $Cu 4s,4p$-orbitals, followed by a secondary 
redistribution
from the $Cu 4s,4p$-orbitals to the $O 2p$-orbitals.
Over all, 0.18 electrons are removed from the 
$Cu 3d_{x^2-y^2}$-orbitals, and put into the $Cu 4s,p$-shell (0.13) and
the $O 2s,p$-shells (2x0.03).
This charge transfer was not detected in the earlier
cluster calculation for the poor $Cu 4s,p$-basis \cite{mest}.
More than 80 percent of this charge transfer arise from the inclusion
of the operators
$n_{i\uparrow}n_{i\downarrow}$ for the $Cu 3d_{x^2-y^2}$-orbitals, the
remaining part stems from the same operators for the $4s,p$-orbitals.

This charge transfer due to atomic correlations is closely related
to the
negative magneto-volume effect known from transition metals as will
be shown next.
The correlation induced charge transfer detected here partially corrects an
inverse exchange induced charge transfer.
The dominant exchange contribution of relevance in
this context 
is from atomic interactions $U_{at}(i)$ and is 
written as $E_{at}(exch)=-\sum_{i} U_{at}(i) n^2_{i\sigma}(\Psi_{HF})$ 
where the sum runs over 
the $i$ atomic orbitals in the unit cell, and $n_{i\sigma}(\Psi_{HF})$ 
indicates the 
HF-occupation
of orbital $i$ per spin. To simplify the discussion,
we restrict to two different atomic orbitals,
and assume that they have the same interaction $U$ but
an occupation $n_{1\sigma}$ very much larger than an occupation 
$n_{2\sigma}$. A
charge transfer z from $n_2$ into $n_1$ leads to the exchange energy gain
$\Delta E(exch)= -2zU(n_{1\sigma}-n_{2\sigma})$. Therefore,
the HF-exchange enhances differences in  charge distributions.
For the case treated here, the HF-exchange is responsible for
a charge transfer
from the little filled $4s,p$-orbitals into the strongly
filled $3d$-orbitals.
Also a trend for a similar
charge transfer from the $4s,p$-orbitals into the strongly filled
$O 2s,p$-orbitals is expected, but should be restricted by 
Hartree (or essentially Madelung) contributions. The latter terms don't
influence intra-atomic charge transfer.
No charge transfer between the $Cu 3d$ and the $O 2p$-orbitals is expected
from atomic exchange as long as these orbitals 
have a similar occupation and not very different atomic interactions.  
When on-site or atomic correlations are included, 
a sizable fraction of the exchange induced 
charge transfer is undone. This is the origin of
the charge redistribution found in the present 
computation.
Another way to undo the exchange induced charge transfer is to turn 
the system magnetic.
This corresponds to a maximal atomic correlation.
Consequently, a charge transfer must come into play when
magnetism in systems with very differently filled subshells is concerned,
as for the case of itinerant ferromagnetism of the $3d$-metals. For $Ni$,
for example one would 
expect a somewhat larger filling of the $3d$-orbitals for
the non magnetic state than for the ferromagnetic state at the cost of the
$4s,p$-occupation, because contrary to the fully magnetic case, the
electrons are not completely correlated in the non magnetic state, as
is well known \cite{soh}.

Such a charge transfer becomes relevant for the magneto-volume effect. 
A theoretical description of
magnetism that is restricted to a particular shell
(like the set of $3d$ orbitals) implies that the antibonding orbitals
are more populated in the magnetic case no matter how strong 
correlations are. In such a case, the
volume always increases with magnetic order. 
When more than one shell is involved, the
just mentioned charge transfer comes into play. For $Ni$, this is  
from the bonding $4s,p$-orbitals
into the (few and antibonding) empty orbitals of the $3d$-shell. It 
implies an negative magneto volume contribution.
$Ni$ actually displays a total negative magnetovolume effect \cite{tt}.
The latter can not be understood within a 
description restricted to the $3d$-orbitals
\cite{soh}, but can only be explained by  a not fully 
screened exchange induced charge transfer. On the HF level,
this exchange induced $3d-4s$-charge transfer 
was first proposed many years ago as the origin of the negative 
magneto-volume effect \cite{le}.

It would be
of interest to find out whether also in the case of the high-$T_c$ compounds
a negative 
magnetovolume effect exists. With the new version of the Crystal program
\cite{crysneu}, which allows for unrestricted HF-calculations, such an
investigation will become feasible from the theoretical side.

When neighbor correlations are included, then an additional charge
transfer of the same magnitude as the one due to on-site correlations
occurs.
It is dominantly from the
$Cu 3d_{x^2-y^2}$-orbitals to the $O 2p_b$-orbitals, and is due to
a particular spin correlation between neighbor $Cu$-sites that will
be discussed later. The longer range contributions that were covered
by the present computations lead to a further but small transfer
of the same kind. A similar but somewhat smaller charge transfer
had been found in the earlier finite cluster calculations \cite{mest}.
It should be noted that this secondary charge transfer is connected
with a small correlation energy gain.
\\

The exact occupation of the $Cu 3d_{x^2-y^2}$ orbitals is of very much
interest. It is directly connected to the measured moment 
of the magnetic ground state
and
plays a crucial role for
models used for the high-$T_c$ compounds.
These usually assume that the occupation
of the $Cu 3d_{x^2-y^2}$-orbital is not very different from 1.0. 
The present calculations indicate that in fact different correlation
mechanisms bring the occupation into this range.
To validify these findings, a short discussion on
the possible deficiencies of the presented computation
shall be given next.

While the result of the HF-calculation can be assumed to be close to
the HF-limit, this does not hold true for the correlation treatment.
A first possible error is connected with the weak correlation approximation. From the
exact treatment of a 
particular on-site correlation plus the connected charge transfer, it can be 
estimated that the on-site
correlation correction of the $Cu 3d_{x^2-y^2}$- occupation is 
overestimated by 10-15 percent. For the charge transfer arising
from longer range correlations, no error estimate can be made. 
Two further corrections are expected. The one is
the influence of shorter than atomic range correlations, the other
is the screening particularly of the $3d$-orbitals due to the $Cu$ core -
and here especially due to the $3s$ and $3p$ orbitals. 
Since
the calculation presented here is the
first ab-initio correlation calculation ever for a
metallic transition-metal compound, no reference results exist.
Even comparable detailed calculations for
small clusters are lacking.
From atomic calculations, some estimate for corrections
can be gained. Such corrections are expressed in terms of
energy differences. In HF-approximation, the excited $3d_94s_2$
state of the $Cu$ atom is 0.4 eV higher than the ground state, while
experimentally the difference amounts to 1.5 eV \cite{at1}.
 A similar correction is found when the ground and excited states of 
the $Ni$-atom
are concerned \cite{at2}. 
This correction is exclusively due to short range correlations,
core polarization effects,
 and relativistic corrections, none of which were included in the presented 
calculation.
The influence of these corrections to the presented results
can be  estimated
 by lowering the diagonal atomic $3d$-energy level by 1 eV. For
the model discussed below,
this
leads to a charge transfer into the $Cu 3d_{x^2-y^2}$-orbitals 
of roughly 0.05.
The lacking short
range correlations alone would have a somewhat larger influence than 
this total shift but are counterbalanced by
relativistic corrections that favor the $4s$-electrons \cite{at3}. 
Adding these corrections, the final estimate for the 
$Cu 3d_{x^2-y^2}$-occupation
amounts to $1.22\pm0.07$. For reasons discussed above, 
it is smaller than the value deduced from the
earlier cluster calculations \cite{mest}.
Future applications will hopefully reduce the uncertainty in the
present LA results.
\\

It is of interest to compare this LA result to an LDA-charge
analysis. 
Table \ref{tab32} also contains LDA results that were obtained for
$YBa_2Cu_3O_{6.5}$ \cite{kgj}.  The latter compound is not 
a so called half-filled system,
and the respective charge analysis represents
a lower limit to the half-filled case.
In the referred publication, only integrated occupations for complete shells
were given. Also, the underlying
charge analysis was performed differently. This might
lead to sizable deviations when the more delocalized orbitals are concerned but
is hoped to lead to comparable results
 for the very localized $3d$-orbitals.
From the global $3d$ occupation, not much can be concluded about
the $Cu 3d_{x^2-y^2}$-occupation. However, an LDA-calculation
performed for the system treated here at half-filling
leads to a  $Cu 3d_{x^2-y^2}$-occupation of 1.55\cite{jepp}, and 
other high-$T_c$ compounds at half-filling are usually
mapped by $Cu 3d_{x^2-y^2}$-occupations of 1.5 \cite{gunnpe}.
These LDA values are very close to the HF-results 
but differ from the correlation result and from the final estimate.

This deviation of the LDA result from the LA occupation is expected to result
from deficiencies of the used homogeneous electron gas
approximation. It is plausible to conclude that 
the specific neighbor $Cu$ spin correlations 
leading to a charge transfer of 0.17 are not at all covered by
the LDA. 
Such a simple connection
can not be made
with respect to the charge transfer caused by on-site exchange terms
and the correlation compensations.
It is known that the LDA is not able to describe anisotropic 
exchange contributions. In particular, the
LDA is
not able to produce the negative magneto-volume effect for Ni \cite{will}.
This indicates that it lacks exchange induced transfer and
partial correlation compensation, but
no large overall error is expected on the atomic scale.
A very rough error
estimate of the  charge distribution due to LDA deficiencies  
on the atomic scale
can be made using an analysis of
LDA results for two-atomic clusters. These indicate that the 
$3d$-orbitals are too attractive in comparison to the $4s$-orbitals.
Expressed in diagonal energies, a correcting shift 
of 1 eV was computed \cite{gujo}. 
Such an atomic LDA correction is similar in size to the joint 
correlation/relativistic correction of the HF-energy differences for the atoms
but has a different
prefactor. It leads to a charge transfer of
0.05 out of the $Cu 3d_{x^2-y^2}$-orbital.

Over all, the LDA seems to overestimate the occupation of
the $Cu 3d_{x^2-y^2}$-orbitals by
roughly 0.3, a large fraction of which is explained. It should
be kept in mind, however, that the atomic orbitals are differently
defined in both methods. This causes some uncertainty.
In the future, it will be possible,
to perform LDA calculations with a new version of the Crystal program
\cite{crysneu}, and to analyze the results by the LA routines, so
that at least this last uncertainty can be removed.
\\

The mutual influence of correlations and charge redistributions
is of relevance to ab-initio methods that try to
adress correlations with Monte Carlo schemes. Variational Monte
Carlo calculations \cite{fah2} as well as diffusion
Monte Carlo calculations (for a review, see \cite{mcrev}) rely
on a good trial state. For the first method,
the charge distribution of the 
trial state is usually frozen to avoid the optimization of
very costly external variational parameters,
while the second method is restricted
by the frozen nodes of the ground state wave function.
The findings of the LA calculation indicate that a better
trial state than the so far always selected
LDA ground state wave function might be needed.

     \subsection{Atomic correlations}     \label{sec:ac}
Next, we discuss
the  individual atomic correlations and their strength.
The average occupation of orbital i in the correlated ground state is
defined as $n_i=\sum_{\sigma}n_{i \sigma}(\Psi_{corr})$.
The
charge fluctuations within orbital i, $\Delta n_i^2$, are given as
\begin{eqnarray}
 \label{eq:densi}
\Delta n_i^2 & = & \langle\Psi_{corr} (\sum_{\sigma}n_{i\sigma})^2\Psi_{corr}
\rangle
-  n_i^2 \nonumber \\
&=&n_i(1-({n_i \over 2}))+2\Delta_i(corr) \\
&=&2(\Delta_i(HF)+\Delta_i(corr)) \\ .
\nonumber
\end{eqnarray}
They are separated into the HF-charge fluctuations, $ \Delta_i(HF)$,
 and the correlation 
corrections $ \Delta_i(corr)$ (see Eq. \ref{fluct}).
The former are defined as the charge 
fluctuations
in a fictitious single-particle state that has the charge distribution of the
correlated ground state. 
These charge fluctuation represent the electronic mobility. 
For a single-particle state, it holds that
\begin{equation}
\Delta_i(HF)=\sum_{j\neq i} P_{ij}^2 \ \ ,
\end{equation}
where $P_{ij}$ represents the density matrix elements per spin to all
other orthogonal atomic orbitals.
The kinetic energy gain due to delocalization of the electrons in this
state is proportional to $\sum_{(j)} P_{ij}$, with the summation $(j)$
restricted to nearest neighbors of i.
Consequently, the reduction of $ \Delta_i(HF)$ due to correlations,
$\Delta_i(corr)$, gives also a rough measure of how much
band energy is lost by the correlations. 

Charge fluctuatons can only be completely frozen out
for half-filling, i.e. for $n_i=1.0$. 
In all other other cases, there is a maximal
correlation reduction, $\Delta_i(corr,max)$, which amounts to
\begin{equation}
\Delta_i(corr,max)=\left \{ \begin{array} {l l}
({n_i \over 2})^2& \mbox{for $n_i<1$}\\
(1-{n_i \over 2})^2& \mbox{for $n_i>1$} \end{array} \right .
\end{equation}
The relative correlation strength $\zeta_i$ is defined as 
$\zeta={\Delta_i(corr) \over \Delta_i(corr,max)}$.

All these quantities are given in table \ref{tab33} for the
final result of the correlated ground state.
As can be seen, correlations are strongest for the $Cu 3d_{x^2-y^2}$-orbital.
Nevertheless, even there, half of the original fluctuations survive,
indicating that the electrons are still very delocalized,
and that a renormalization of the effective mass due to atomic
correlations of no more than 30 percent is to be expected. 
Nevertheless, 70 percent of
the possible reductions are realized. Next in strength are the correlations
on the $O 2p_b$-orbitals. Also here,
the correlation strength is 0.7 although the reductions amount
to only 20 percent of all fluctuation in this orbital.
The correlations in all other orbitals are weak. This even holds true for
the $O_{2s}$-orbital.

The correlation strength strongly depends on the included 
correlations.
When restricting to on-site 
correlations, the $Cu 3d_{x^2-y^2}$-occupation is 1.33. Then,
it holds that $\Delta_i(corr)=-0.096$, which represents a 85 percent reduction.
Freezing the $Cu 3d_{x^2-y^2}$-charge at the HF, value, i.e. close to the
value of the $O 2p_b$-charge leads to a correlation strength of more than 
0.90. This will be explained later when analyzing these correlation functions
in the context of model interactions. 

A set of trial variational calculations restricted
to individual correlation operators was also performed. This was done
to control the validity of the variational expansion.
When comparing these variational results to the
variational expansion results, it was found that the correlations
obtained by the expansion calculation,
 were overestimated by 10-15 percent for the $Cu 3d_{x^2-y^2}$-orbitals,
but less than 5 percent for all other on-site correlations. This small
corrections indicate that the expansion is fully able to cover these
correlations. Consequently, there is no evidence that
correlations on the atomic scale are too strong for a weak correlation
expansion treatment.

There are additional corrections expected from the omitted correlations.
From earlier calculations for other systems, it was found that the
longer range correlations that were neglected here have
no influence on the atomic  correlation functions (for a detailed explanation,
see eg \cite{8}). The short range correlations omitted here, however, led to
a reduction of the atomic correlation corrections
 for the $2s,p$-orbitals by $10 \pm 5$
percent \cite{8}. Consequently, they are expected to lead to a somewhat
larger reduction for the $3d$-orbitals which are characterized by a somewhat higher average
density. When added, a reduced correlation strength of
$0.53 \pm 0.04$ instead of 0.7 is expected for the $Cu
3d_{x^2-y^2}$-orbitals, when  an occupation of 1.17 is assumed. 
Such a correlation correction  is not terribly large but is  rather
similar to the correlation strength obtained for the transition
metals from model calculations \cite{tra2}.

This correlation strength obtained for the metastable haf-filled state
should not strongly change when electrons are removed from the planes.
Actually, for such a case, the $Cu 3d_{x^2-y^2}$-occupation 
comes closer to 1.0, and $\Delta_i(corr)$ is expected to increase while 
$\zeta_i$ decreases. Therefore,  the results obtained for the
half-filled case are expected to be representative for relevant dopings.

Complete magnetic
order implies a
correlation strength of  1.0. Consequently, we would expect a
certain  additional
charge transfer $3d-4s,p$ with magnetization. This might even
lead to a negative magnetovolume effect.

     \subsection{Spin correlations}     \label{sec:sp}
Of particular interest are the spin correlation functions
$S(i,j)$ between the different atomic orbitals $(i,j)$
, defined as
\begin{equation}
S(i,j)=\langle\Psi_{corr} |\vec{s}_i \vec{s}_j| \Psi_{corr}\rangle \ \ .
\end{equation}
 For the on site terms, it holds in
general that
$S(i,i) ={3 \over 2}(\Delta_i(HF)- \Delta_i(corr))$.
For the SCF- ground state, these expectation values don't vanish, and
represent
the auto-correlations of the electrons. They 
are small  except for the on site terms and the neighbor $Cu-O$
contributions.
For the correlated ground state,
it turned out that only spin
correlations inbetween the $Cu3d_{x^2-y^2}$-orbitals on the different
atoms are relevant. This is in contrast to
earlier calculations for small clusters. 
There, longer range spin density wave like 
correlations
occured at which the  $O 2p_b$-orbitals participated \cite{mest}.
Apparently, these small clusters rather represented
a 1-dimensional 
chain than the intended plane. 

Table \ref{tabss} 
displays the correlations $S(i,j)$, where the indices $(i,j)$ are
restricted to $Cu3d_{x^2-y^2}$-orbitals. 
These correlation functions are computed incrementally.
In a first step, only on-site correlations are included, and only the
on-site function is given, then neighbor interactions plus the neighbor
functions are successively included.
The enhancement of the on-site terms with inclusion of
longer range correlations is due to the enlarged charge 
transfer.  

The antiferromagnetic $Cu$-neighbor spin 
correlations are large
and are connected with a sizable charge transfer from $Cu$ to $O$
that was discussed above.
The coupling of spin correlations 
to a charge transfer arises because spin correlations are maximal
for a $Cu 3d_{x^2-y^2}$ occupation of 1.
In more detail, the
quantity most relevant for the spin correlation strength
is the expecation value  $\langle \vec{s}_i \vec{s}_j H \rangle$,
where $i,j$ represent neighbor $Cu 3d_{x^2-y^2}$ orbitals.
The dominant contributions from the Hamiltonian are the
on-site interactions
on $Cu$, $U_{3d}$, and on $O$, $U_{2p}$. When restricting
to these interaction, and to the largest density matrix elements $P_{ij}$,
the above expectation value is given by
\begin{equation}
\langle \vec{s_i} \vec{s_j} H \rangle ={3 \over2} 
(2U_{3d_{x^2-y^2}}n_{i\sigma}(1-n_{i\sigma})P_2^2-U_{2p}P_1^4) \   \   .
\label{sisj}
\end{equation}
Here, $n_{i\sigma}$ represents 
the SCF-occupation of the
$Cu$-orbital per spin, and $P_1$ the neighbor $Cu-O$ density matrix element 
and $P_2$ the
neighbor $Cu-Cu$ density matrix element. It holds that 
$n_{i\sigma} > |P_1| > |P_2|$ for the relevant range
of occupations. 
For the half-filled SCF-ground state, the first term of 
Eq. \ref{sisj} is one order of magnitude
larger than the second.
The correlation induced charge transfer results probably because 
the matrix element  $\langle O_{\nu} H \rangle$
is enhanced by a charge transfer from the $Cu$-orbital to the $O$-orbital. 
Due to such a transfer,
the first part increases, while the
second part strongly decreases. The charge transfer stops when the
$Cu 3d_{x^2-y^2}$-occupation reaches 1.

The ab-initio results show that
the charge transfer due to the
secondary spin correlation is large. This suggests that 
the ground state must also be relatively instable with respect to any
other external disturbance that profits from such a
charge transfer.
May be this instability also contributes to
the lattice instability, and in particular to the large buckling 
found for this compound \cite{and2}.
In view of this sensitivity, it is even more astonishing that the 
very much larger on-site correlations did
not lead to a sizable $Cu-O$-charge transfer. 

These neighbor correlations are quite strong, stronger
than needed to counterbalance the change in the wave function due to
on site correlations. 
For a singlet state $\Psi$ it always holds
that $\sum_{ij}\langle \Psi | \vec{s_{i}}\vec{s_{j}} | \Psi
\rangle=0$. For the SCF-ground state, the on-site terms are
counterbalanced in part by antiferromagnetic
neighbor $Cu-O$ spin correlations, while the
remaining correction is rather long range. When correlations up
to
neighbor $Cu$ atoms are included, then the neighbor
$Cu-Cu$ correlations alone more than   
counterbalance the correlation enhanced on-site terms.
This can be viewed as
a quite sizable attraction of electrons of different spin on
neighbor sites even in the absence of
longer range antiferromagnetic order.
In fact, it will be later demonstrated that these correlations depend
little on doping.
\\

While the nearest neighbor correlations represented in table \ref{tabss}
are
converged with respect to the treated cluster sizes (as
long as no longer range correlation operators are added), 
this does not quite hold true for the
second nearest neighbor terms. Here, another ten
percent might be obtained from more extended cluster contributions.
Third nearest neighbors when added
are not converged at all. Here a renomalization of at least a factor of two  
is expected from larger clusters. 

More extended clusters and longer range spin correlations
were not covered since alredy
the results obtained so far led to spin correlation corrections 
that turn too large.
This indicates the proximity to the antiferromagnetic
instability. In some individual cluster calculations,
already next nearest neighbor contributions turned out to be almost as large
as nearest neighbor terms. 
Nevertheless, for the included clusters the expected
computation breakdown was not yet seen. Such a 
breakdown must arise as soon as a cluster size is reached where in
HF-approximation a broken symmetry ground state is preferred. 
Consequently it is concluded that even for the
half-filled case antiferromagnetic order parameter
fluctuations need to extend over domains that are 
larger than the cluster explicitly included in the computations.

The longer range correlations contribute differently from
the nearest neighbor correlations. The direct coupling matrix elements
$\langle O_{\nu} H \rangle$ are negligible. However, the longer range
operators
are more sensitive to electrons very close to the Fermi surface.
Consequently, spin correlations may form that are dominated by
the band electrons very close to the Fermi energy.
\\

It is of interest to compare these correlation functions with the ones
of a two-dimensional Heisenberg model \cite{manu}. For the latter, 
the on-site and neighbor 
correlation functions are also given in table \ref{tabss} 
together with the long range
limit. It can be seen that the short range correlation 
functions of the Heisenberg
model are larger than the ones of the real system.
This is because in the real system the
$Cu 3d_{x^2-y^2}$-orbitals are more than half-filled, and because they
are not  perfectly  correlated.
On the other hand, the short range correlations of the real system
extend already beyound the long range correlation pattern 
of the Heisenberg model.
The limiting correlation function in the magnetically ordered
high-$T_c$ materials is typically $\lim_{\nu \rightarrow \infty} 
|S(i,i+\nu)|=0.06$ \cite{jurg}.
This is well below the short range correlation functions obtained from the
present calculation.

     \section{Determination of a model Hamiltonian }  
                              \label{sec:mod}

The ab-initio results of the LA provide sufficient information
to unequivocally determine a model Hamiltonian. Such a condensation
of ab-initio results  to a model serves
multiple purposes. 
One of them is to forward ab-initio informations 
to computations that can no more be performed on an
ab-initio level but only for a model. In the following,  this applies
to the computation of the doping dependency of 
the properties discussed above that were calculated 
for half-filling. Due to the restrictions
of the Crystal program, such ab-initio calculations would
become costly. 
Another purpose
is that information about particular correlations can be represented
in form of effective interactions. 
Usually, experiments are fitted
by models which are represented by the adapted interactions, but
no correlation function for the model is computed.
Such a connection will facilitate comparisons, also for
differing systems.
For the case of the
high-$T_c$ compounds, finally, the explicit treatment of correlations
was so far restricted to models.
Therefore, it is of interest to see, how well such models match 
the ab-initio findings.

The determination of a model from the ab-initio data separates into two steps.
The first is the choice of the model space, i.e. which orbitals to
include,  and the computation of the relevant single-particle
Hamiltonian.
In the second step, the effective interactions are computed.

     \subsection{Relevant single-particle space and 
single-particle Hamiltonian }  

                              \label{sec:mod1}
The information provided by the ab-initio calculation
concerns atomic orbital degrees of freedom, but
so far, 
no more delocalized degrees of 
freedom
like Wannier orbitals extending over sets of atoms. Models built
from delocalized orbitals can not be directly compared to the
ab-initio data but would need to be derived in a second
step from models based on atomic orbitals.

The
smallest model used for the high-$T_c$ compounds
based on atomic degrees of freedom
is a 3-band model, containing as atomic orbitals
the $Cu 3d_{x^2-y^2}$ and the $O 2p_b$ orbitals.
However, from the ab-initio charge analysis 
(see table \ref{tab32}), one notes non negligible fillings of $Cu 4s,p$ and
even deviation from complete filling for the
$O 2s$-orbitals. It is also known that the $Cu 4s$ orbitals
contribute actively to the band structure of the most relevant 
half-filled band. This was
worked out before from LDA calculations \cite{and1}.
As a compromise, the selected
model space for the present application is chosen to consist of
$Cu 3d_{x^2-y^2}$ and $4s$ 
and the $O 2p_b$ orbitals. This model is also selected because 
an LDA equivalent exists \cite{and1}.

Note that these orbitals can not
be seen as a 
perfect representation of the corresponding ab-initio 
orbitals. When taking the charge distribution for the SCF-ground
state from table \ref{tab32}, then the orbitals included in the model 
represent 4.84 electron per unit cell for the SCF-ground state and 4.89
for the correlated ground state instead of 5.0 as they do for the 
half-filled band case in the model.
Consequently, there can be no perfect agreement between such a 4-band model
and the real system. 

Instead, the most relevant properties are
to be matched. Here the following properties are selected.
The first is the half-filling of the uppermost
band. This fixes the model charge at 5 electrons per unit cell.
The second is the exact
occupation of the $Cu 3d_{x^2-y^2}$ orbital obtained from the respective
ab-initio calculation. Since the influence of particular correlations
will be investigated, also ab-initio calculations with only partial
inclusion of correlations and varying  $Cu 3d_{x^2-y^2}$-occupation 
will be fitted.
The third is the form of the Fermi surface. The model Fermi surface
shall match the ab-initio Fermi surface. This
is important when longer range correlations are concerned, and
puts restrictive bounds on the $Cu 4s$-occupation.
A model $Cu 4s$-occupation taken from the
ab-initio result would lead to a Fermi surface that deviates 
too strongly from nesting. Consequently, the $Cu 4s$ model occupation is 
set to 0.25. The omitted $Cu 4s$-charge in the ab-initio calculation
apparently stems from bands omitted in the model.
Fixing the two other occupations freezes the $O 2p$-occpuation.
It turns out that the deviation of the latter 
from complete filling is only half as large for
the model as for the true ground state, indicating the bias and the
limits of the 4-band model.

Having determined the model  occupations for a particular ab-initio fit 
implicitly defines the 
diagonal or crystal field terms $e_i$ of the model.
These $e_i$ contain exchange and correlation contributions of the
omitted degrees of freedom as well as exchange contributions
due to the added on site interactions of the model. Consequently,
they differ for every fit.
Each time, they are determined selfconsistently
for the model calculation so that the intended charge 
distribution is obtained.

The second set of parameters describes the delocalization of
the electrons. It consists of the hopping terms. Here, it is
assumed that the omitted external degrees of freedom
have no influence on these terms. Since only on site interactions
will be included, no nonlocal exchange contributions arise
in the model. The hopping terms therefore represent the
non diagonal Hartree or alternatively LDA matrix elements.
There are only two relevant hopping parameters for this 4-band model.
From the LDA-fit, it holds for the $3d,2p$ hopping,
$t_{dp}=-1.6$ev and the $4s,2p$-hopping, $t_{sp}=-2.3$eV \cite{and1}.
At present,
these hopping elements can not be directly computed by the LA 
since Crystal92 \cite{crys}
does not separate kinetic energy (plus Hartree)
terms from exchange terms. However, an estimate can be made.
The Fock matrix elements,
$f_{ij}=t_{ij}+V_{ij}(exch)$, can be computed. For them,
it is found that $f_{dp}=-2.8$eV and  $f_{sp}=-4.2$eV. 
When approximately correcting for the exchange using the relation 
$V_{ij}(exch) \simeq -{1\over{|\vec{R_i}-\vec{R_j}|}}P_{ij}$,
values of $t_{dp}=-1.7$eV and  $t_{sp}=-2.9$eV result. 
The $t_{pd}$ term equals the LDA value. This
indicates that in the future hopping terms between orthogonalized
atomic orbitals for models can be directly computed by the LA.
On the other hand, the $t_{sp}$ from the-LDA fit
is 20 percent smaller. This difference probably arises because
the LDA-fit accounts
for the omission of
the $4p$-orbitals which otherwise would change
the higher energy band, while the estimates from the LA
represent the original bare hopping. 
In the following, the LDA values for the hopping are taken.

The charge distribution of the original  single-particle LDA model 
Hamiltonian
(with the LDA values for the $e_i$ \cite{and1})
amounts to a  $Cu 4s$-occupation of 0.3, also
strongly reduced due to the need for an adequate Fermi surface.
The
$Cu 3d_{x^2-y^2}$-occupation for this model is 1.4, and apparently
represents the $Cu 3d_{x^2-y^2}$-contributions of the four included
bands.

     \subsection{Interaction terms}  
                              \label{sec:mod2}

For the determination of interaction terms in the model,
correlation functions are available. In the model, only atomic or on-side
interactions are included. These are the diagonal interactions for
electrons in the same orbitals, 
$U_{3d}, U_{4s}, U_{2p}$
and the interaction between the $Cu 4s$ and $Cu 3d_{x^2-y^2}$-orbitals
on the same atom,  $U_{4s,3d}$.
These interactions are fitted on a one to one basis with the
help of
the corresponding on site correlation functions.

The effective local model interactions $U_i$ are indirectly
generated from the long range Coulomb interaction that prevails
in the ab-initio calculation. In this
process, different kinds of rescaling occur. One rescaling process is called
folding. It is a reduction that is not connected to screening.
When reducing a single atomic fluctuation then not the original
atomic interaction is measured but the difference between
this interaction and the residual interaction
of the electrons shifted in the process. For an almost empty band, the
residual interaction is zero while in zeroth order, for a half-filled
band, it is the neighbor interaction. A more detailled discussion
was made in Ref. \cite{8}. 
Another rescaling is due to
screening effects, and here two sources exist.
The one are the degrees of freedom not included in the model, and the
other are correlations also present in the model but not activated
for  the lack of a longer range model interaction.

In the following we will add correlations stepwise
to the ab-initio calculation, and will stepwise interpret the ab-initio
result in terms of modified model interactions. This way, the derivation
of a local $U$ from the bare Coulomb interaction can be quantitatively
understood. The end product are local model interactions that 
represent the true ground state.
The actual computation procedure for the model is as follows. 
The ground state for
the single particle part of the model Hamiltonian is easily obtained.
It is used as an input into the LA program like in the ab-initio case
the Crystal92 results.
The correlation calculation is then performed by the LA program package,
but with the interactions reduced to the model interactions.
In principle, for the model, a set of calculations for different clusters would
be necessary as in the ab-initio case before. In reality, the fit was
performed by matching only correlations in one, namely the largest cluster
to the corresponding ab-initio results.
This cluster consists of
five active $Cu$ atoms and four active $O$ atoms.
Every model single-particle calculation
and subsequent correlation calculation are embedded into a 
self consistent cycle in which the respective single-particle energies
$e_i$ of the model are
fixed so that the charge distribution of the particular
correlated state
of the model matches the ab-initio counterpart.

When comparing model and ab-initio  correlations, 
an additional constraint needs to
be taken into account. It is that correlation functions can only be
directly compared when the respective
occupations of the ab-initio calculation and
the model are identical. A similar constraint arises from the folding
effect on the value of $U$. This is also strongly occupation dependent.
While this poses no problem for the $Cu 3d_{x^2-y^2}$ orbital, it
involves the other two whose charges don't match.
Therefore, interaction terms for these orbitals are
determined in an intermediate step in which the
model charge balance
between the $Cu 4s$-orbitals and the $O 2p$-orbitals is shifted so
that it agrees to the ab-initio result for the momentarily treated atom.
Fortunately, the different interactions don't influence each other much, so
that no sizable ambiguity arises from this procedure. A test can
be made by recomputing
$U_{3d}$ for the different choices. $U_{3d}$
varies by less then 10 percent. It is largest for the highest $4s$
occupation because then the
 $4s,3d$ screening explicitly handled by the model itself is largest.
For the other occupations,
larger fractions of this screening are mapped by a reduced interaction 
parameter   $U_{3d}$.

Sets of parallel calculations were performed in the following steps.
First, correlations were introduced on a single atom only, once for
$Cu$ and once for $O$. In both cases atomic charge transfers were 
alternatively allowed or
blocked to investigate their effect on $U$.
 The corresponding
ab-initio calculation was also restricted to correlations
on single atoms only.
This leads to unscreened effective  interactions $U$ when only
single correlation operators each are included.
These 
interactions $U$ although unscreened are
folded from the Coulomb interaction and do no more  represent
the original atomic interaction matrix elements. 
With charge transfer excluded, the results
are given in the first line of table \ref{tabmod1}. Although folded,
the $U_{3d}$ is not very far from the atomic interaction matrix element which
is expected to be $U_{atom}\simeq 25$eV. The folded interaction $U_{2p}$
on $O$
on the other hand is quite small. It compares to similar interactions
obtained before for the atomic orbitals on $C$ compounds \cite{8}.

The values for $U_{4s}$ and $U_{4s,3d}$ are very small.
The folding effect resulting from neighbor interaction contributions
is expected to be relatively largest for the latter terms because
the $4s$-orbital is most extended.
Also it should be remembered
that the model contains effective $3d-4s$
hopping terms that are 20 percent smaller than 
the original values. For weak correlations it holds that correlations
scale $\simeq {U \over t}$.
Correcting for this renomalization would lead to an 25 percent
enhancement of the $U_{4s}$. It shall be finally reminded that part
of the ab-initio $4s$ occupation apparently stems from other bands
and might be involved in hopping processes in the ab-initio calculation
that are not represented in the model. Taking this into account should
lead to an additional enhancement of $U_{4s}$. For comparison, the value for
the
effective local interaction in another system with $4s,4p$ electrons,
namely $Ge$, is 3.1eV \cite{diam}, which is not too different from
such a rescaled value.

The direct correlation between the 4s and the 3d orbital on the $Cu$ site 
when added has
only a very small  influence on the value of $U_{3d}$, 
and the value of
the $U_{4s}$ and$U_{4s,3d}$ is not very relevant for $U_{3d}$.
This is unexpected. 
Similar interactions were important in earlier applications when the 
screening between
$\pi$ and $\sigma$ electrons in $C$ isomorphs or organic compounds was 
concerned \cite{8}. In these cases electrons in half-filled bands were 
screened by electrons in wider bands that were also
half-filled. 

Next, local charge transfers are allowed that arise due to correlations.
 For the single $Cu$-site, this is
a charge transfer from the $3d$ into the $4s$
orbitals. It leads to a strong reduction for $U_{3d}$ (see the
second line in table \ref{tabmod1}). 
Partly, this reduction originates from the
change in occupation itself because  the folding reduction of the original
atomic interaction due to longer range
 Coulomb terms is largest for half-filled
atomic orbitals \cite{8}, and the change in $Cu 3d$ occupation is toward
half-filling.
The dominant contribution is from the $4s,3d$-screening.
Such a screening was proposed by Herring long time ago \cite{herr}.
Astonishingly, it comes into play with the help of a charge transfer
when starting from the uncorrelated ground state, and the 
major role of the $4s$
orbitals is to represent a reservoir of states. 
This is interpreted in the following way.
In HF-approximation, the fluctuations are somewhat reduced by a too large
occupation as compared to the one of the true ground state
(see the discussion of the magneto volume effect in section \ref{sec:ac}).
The correct occupation is then obtained by the added
single-particle correlation operators. This way 
less fluctuations are induced than would arise
within a uncorrelated ground state that has the correct occupation.
There are residual fluctuations, part of them from the
original uncorrelated ground state, and part from the charge
transfer due to single-particle
correlation operators. These fluctuations are reduced with the help
of the two-particle operators. 
In the model, a sizable part of the screening electrons is no more included.
When fixing the $3d$-occupation at the correct value, then very much 
larger fluctuations arise for the model single particle ground state
than for the original ab-initio uncorrelated ground state.
With the original $U_{3d}$, a suppression
of fluctuations for the model would result
that is very much larger than the suppresion
of charge fluctuations due to two-particle correlation operators in the
ab-initio case. Consequently, a false description of the 
$3d$ correlation corrections
would be made.
When the reduction of fluctuations is adjusted to the known
correction a very much reduced effective
interaction $U_{3d}$ is found.
 
This scenario for the $4s,3d$-screening is
very different from the previously treated cases.
As just mentioned, the screening of the electrons in the 
half-filled $\pi$-bands in 
organic systems or
$C$-isomorphs due to the electrons in the  half-filled $\sigma$ bands
was not at all connected with a charge 
transfer but originated
solely from two-particle correlation operators, i.e.
can be seen as a kind of classical screening.

For $O$, no on-site charge transfer effect did occur. It is plausible that the
only possible charge transfer which is
with the not completely filled $2s$-orbitals is marginal.
In a subsequent calculation, all on-site correlations in the cluster
were treated at once, and also charge transfer between the atoms was
allowed. The result is displayed in the third line of Table \ref{tabmod1}.
There is an additional small charge transfer out of the $Cu 3d$-orbitals,
leading to a small further reduction of $U_{3d}$. From this term on,
the $U_{3d}$ obtained for the minimal $4s$ occupation are given.
Changing to this reduced value of the $4s$ occupation contributes
a reduction of $U_{3d}$ of 0.8 eV. This reduction arises because the
residual $4s$ occupation is connected with a smaller screening. 
There is also a
large charge transfer into the $O 2p$ orbitals. This causes a reduction
of the folding effect and a sizable enhancment of $U_{2p}$. Remember that
on this level no other but on-site correlations were
included in the ab-initio treatment. In particular the long range part of the 
Coulomb interaction was not screened at all. 
When nn $Cu-O$ correlations are also included (line 4 of Table \ref{tabmod1}),
both model interactions are further reduced. The effect is larger for
$U_{3d}$, where a charge transfer and screening come together,
and smaller for $U_{2p}$, where a screening gain is reduced by
another  enhancment due to the inverse  charge transfer.
A sizable further reduction of $U_{3d}$ occurs when all corrrelations
in the cluster are included. This is mostly due to the $Cu-Cu$ spin
correlations and the $3d,2p$ charge transfer caused by them. 
For $U_{2p}$,
no value was computed for this case. A computation of $U_{2p}$ requires
a strong charging of the $4s$ orbital. In this particular case, it is
so large that the $3d$ orbitals are overscreened. This means that
even for sizably enhanced $U_{3d}$, too small correlation corrections
of the $3d$ charge fluctuations were obtained. When keeping the
original $U_{3d}$ obtained for the other distributions, a value
of $U_{2p}\simeq 9$eV was obtained. This represents a further enhancment due
to the charge transfer into the $O 2p$ orbitals.  

The values for the model interaction parameters obtained this way 
can be seen as 
upper limits for the true model parameters. This is because only
specific correlation corrections were so far included. Part of the excluded
correlation
corrections can be estimated. 
This is first the effect of the
very short range intra-atomic correlations. 
Their inclusion led to a
 reduction of the $U$ by 10 percent or 0.5 eV for the $C$-isomorphs 
\cite{8}.
A similar reduction is expected for $O$. For $Cu$ a somewhat larger
reduction is expected due to the higher density and due to the 
screening influence of the filled $3s$ and $3p$ shells. 
This is next the omitted long range and polarization 
corrections. Such corrections are irrelevant for half-filling. 
Consequently, they should have no large effect on the $U_{3d}$. However,
they should at least reduce a large fraction of the enhancement of $U_{2p}$ with
charging. 
The resulting estimate of the model parameters is given
in the fifth line of Table \ref{tabmod1},
togehter with a rough error estimate.
Note that this is the first calculation of its kind for
an ionic compound as well as for a metallic compound
containing transition-metal atoms. Future
applications will certainly reduce the uncertainties of the error estimates.
\\

Values for the effective atomic interactions were so far computed
by LDA frozen charge calculations. From these, an effective local
interaction $U$ is obtained that does not distinguish between
different $d$-orbitals and the total angular momentum of the
atomic charge. 
In a second step, higher order Slater parameters
were added that are taken from experiments on atoms. An
introduction is given in Ref. \cite{mam}. 
Table \ref{tabmod1} contains an average over 
LDA results \cite{mms,sj,hyb,mam} for the resulting 
diagonal interaction of the $3d_{x^2-y^2}$
orbitals, $U_{3d}$, which here is identical to the diagonal
interactions of the other $3d$-orbitals.
Partially, those calculations also
contained results for neighbor Coulomb interactions $V$ (typically 1 eV
or smaller). In such cases,
the values presented in table \ref{tabmod1} are
the differences $U_{3d}-V$. 
The LDA interaction is considerably larger
than the one found from the LA calculation,
and would result in too large correlations if used for the model.

Note that the definition of $U$ is totally different in the two
approaches. The one (LDA) freezes charges and does not care
for their dynamics, i.e. whether they are essentially localized
or whether they are delocalized. Also, 
only the nearest neighbor environment matters. So for $CuO$, almost the
same interaction is obtained as for the high-$T_c$ materials \cite{mam}.
The other method (LA) maps all
particular correlation effects even due to longer range interactions
of the delocalized electrons into an effective folded local interaction
$U$. When looking for the derivation of the LA value, then it is
seen that for the considered system,
a very peculiar $Cu$ neighbor interaction (or spin correlation)
leads to a reduction of 2.5eV below the LDA-value. Such an
effect is $3d_{x^2-y^2}$ specific and would not be expected to play a role
for the other $3d$-orbitals. It would also not be expected for $CuO$
for a lack of neighbor $Cu$ coupling.
For the latter compound, the $U_{3d}$ of the LA are expected to be
in the range of the LDA values. 

The LDA results were used to explain photoemission experiments
for different transition-metal compounds. Photoemission
spectra calculated for an Anderson impurity model with the computed LDA
values of $U_{3d}$=9.5 eV \cite{gunnpe} led to very good agreement
with experiment for $CuO$ and (with a particular exception) for $Nd_2CuO_4$.
A model when mapped to the same experiment led to a value of
 $U_{3d}$=8.4 eV for $CuO$ \cite{ets}. This demonstrates that the LDA
values are the correct values for the interactions among the
completely localized $3d$-orbitals. However, this can not be
taken as evidence for the correctness of the value for the
 $3d_{x^2-y^2}$ orbitals. A modification of the latter interaction towards
the LA result would probably not change the computed spectra
very much. 
There is a small deficiency for $Nd_2CuO_4$, though,
when fitted to the LDA values. This is
the existence of a local singlet peak at the upper band edge
in the calculation, which   also shows up in calculation and experiment for
$CuO$, but not in the experiment
for $Nd_2CuO_4$. 
 It is a valid
speculation whether already the
reduced $3d_{x^2-y^2}$ $U_{3d}$ originating from the
effective neighbor $Cu$-correlations in the plane
would remove this deficiency.

The $U_{2p}$ interaction
of the LA is in qualitative agreement with earlier values found
for the $2s,2p$-interactions in diamond (7.2 eV \cite{diam}) but
 larger than the LDA estimate.
There exists a spectroscopical fit for $U_{2p}$=5.5eV \cite{ed}.
\\

The presented difference between the LA values and the LDA values
of $U_{3d}$ matches the difference between LDA and experiment found
earlier for the transition metals.
For the transition metals, the $U_{3d}$ of the LDA are
apparently independent of band filling \cite{gunnu1,gunnu2} 
while the experimentally
needed quantities are strongly filling dependent and considerably
smaller - except for the
completely filled $3d$-band limit \cite{tra2,soh,8,saw}.
It was proposed to resolve this
discrepancy and the filling dependence of $U_{3d}$ by a not fully screened
neighbor interaction $V$. This would explain the LDA deviations and
the filling dependency 
of $U_{3d}$ as a folding effect \cite{8}.

Calculations with the LA are now feasible for transition metals.
The results obtained here give hope that
from such calulcations, apropriate values for the $U_{3d}$ of the
transition metals can be obtained. 
For comparison, the value of $U_{3d}$ needed for $Ni$ is 4.7eV,
and not very much lower than the final LA estimate for $Cu$ in $SrCuO_2$.

     \section{Model interaction and spin correlations}  
                              \label{sec:mod3}

The model interactions derived in the last section were optimized with 
respect to charge distribution and on-site correlations. 
Next, we will control how well this
model is also able to reproduce the most interesting longer range 
correlation
features, namely the spin correlations between different $Cu 3d$-orbitals.

For the 5 $Cu$-cluster, the results of the
model calculation with $U_{3d}=6.3$eV are compared to
the corresponding ab-initio values
in table \ref{tabmod2}. As can be seen, the model
neighbor spin correlations are only half of the ab-initio values, and the 
2nd neighbor terms are reduced to one sixth. 
Consequently, the 4-band model with only on-site interactions included 
can not consistently reproduce on-site correlations and neighbor
spin correlations.
Ab-initio neighbor spin
correlation functions are only matched by a
$U_{3d}$ that is 50 percent to large, and even then,
the deficiency of the second neighbor
correlation functions is not completely removed (see table \ref{tabmod2}).

Part of these discrepancies can be understood by considering the
$U$ dependence of the individual
correlation functions calculated by means of the
weak-correlation expansion. Here, the proximity to the magnetic instability
plays a particular role.
For the model (and for the
ab-initio calculation), the interaction enters in two quantities, namely
$\langle OH\rangle$ and $\langle OHO \rangle_c$. If the 
$\langle OHO \rangle_c$ were not depending on $U$,  then the variational
parameters (and the correlation functions up to saturation)
would rise linearly with interaction strength. 
The model results are different and indicate that the interaction dependency
of the second terms must come into play. 
This holds particularly true for the
longer range correlations who rise
very much more than linearly with interaction. 
This anomalous $U$ dependency
can only be understood by the proximity of the
magnetic phase. As discussed before, the terms
 $\langle OHO \rangle_c$ represent the 
two particle excitation energies. Close to a magnetic instability,
these might become very much smaller and tend to zero, leading to
an anomalous $U$ dependence of the correlation parameters.
Apparently, this applies to the model.

This interpretation also explains why the longer range spin correlations in
the model are relatively weaker than for the ab-initio case.
The diagonal terms  of $\langle OHO \rangle_c$  represent energy differences of
bare excitations out of the SCF-ground state wave function.
This means that in the ab-initio calculation,
for these matrix elements the
uncorrelated or HF susceptibility enters. In the model however,
the calculations were not performed with bare but with screened interaction
parameters. This means that for the longer range spin correlations the energy
difference of bare excitations is computed with screened interactions
and therefore contains correlations to some extend, in contrast to
the ab-initio calculation.
Consequently, the
model result for the longer range spin correlations might be more adequate than
the ab-initio result and might even indicate by which amount the ab-initio
results need to be corrected.
With only
second or third neighbor correlations included, the model is still far from
instability, in contrast to the conclusion derived from
 the ab-initio calculation.
A magnetic instability might only occur when considerably longer-range
magnetic correlations are added. It can even not be excluded that
the charge transfer connected with magnetic correlations
causes a first order phase 
transition, and that no divergency
of the long-range correlations can be detected in the
metastable state without broken symmetry.

Nearest-neighbor spin correlations don't display
such a strong $U$-dependence. While it can not be excluded that
the difference between ab-initio and model results might also
originate from the overestimated Stoner enhancement, there is
another deficiency of the model that points to a different source.
The neighbor spin correlations
are connected to an explicit $Cu 3d$-$O 2p$ charge transfer
that is by almost one order of magnitude smaller
in the model than in the ab-initio calculation. 
A reason for this may be that the $Cu 4s,p$ degrees of freedom
are mostly removed  from the model and only indirectly included
in the form of reduced on-site interactions. 
It might well be that these omitted degrees of freedom
contribute more actively to
the neighbor-$Cu$-spin correlations with the help of
an induced magnetic exchange interaction between neighbor $Cu$
sites.
Another reason for the discrepancy between ab-initio and model results
might be that the $O$ occupation in the model is considerably larger 
than in the ab-initio case, and might also reduce the charge transfer.

Apparently,
the 4-band model with on-site
interactions only is not quite
adequate to deal with the most interesting outcome of the ab-initio
calculations, namely the anormalous neighbor spin correlations.
\\

The Stoner enhancement in the longer-range spin correlation 
functions of the model calculation
 of the 4 (or 3) band model and also apparently in the 
ab-initio calculation is very different from
results expected for a single band Hubbard model. When using
a weak correlation expansion, then it is well known that for
the one-dimensional case all interaction contributions in the
terms $\langle O H O \rangle_c$ drop out. This means a 
linear rise of all correlations with $U$ but no Stoner
enhancement. For the two-dimensional model with
perfect nesting, similar results are expected. It can not be ruled
out that closed loop terms lead to interaction contributions in the
$\langle O H O \rangle_c$ for the non nesting case
 but such terms are not yet present in the considered 
5 atom cluster. 
This indicates that the magnetism in the real system is essentially of 
itinerant or spin-density-wave nature ( although strongly enhanced by the
almost perfect nesting), and that a simple single-band Hubbard model might
not to be the correct approximate description.

     \section{The model away from half-filling}  
                              \label{sec:mod4}

The computations performed so far were restricted to the so called
half-filled band case. As mentioned before, the program Crystal92 can only
be used for an integer number of electrons per
unit cell. There is no such restriction
for the LA program package. Consequently, the model calculations can easily
be extended to partial fillings. 
For simplicity, the model SCF calculations were not repeated for differing
fillings but the single-particle Hamiltonian at half-filling was frozen
in, and only the Fermi energy was shifted.
This approximate treatment
seems justified because contributions relevant for charge redistribution
like the long-range Madelung terms are not included in the
simple on-site interaction model.

Of interest is the dependency of the neighbor and of the longer range 
spin correlations on band filling. Fig. \ref{fig10} displays 
the nearest-neighbor ($\nu=1$) and next-nearest-neighbor ($\nu=2$)
$Cu$-spin correlations as a function of the occupation of the 
uppermost band ($n_B$). These correlation functions were
taken from calculations for a five $Cu$ atom cluster again. Corections
towards the full results are typically 20 percent for the nearest-neighbor
terms and more than 100 percent for the second-nearest-neighbor contributions.
Two values for the interaction parameter $U_{3d}$ were taken, namely the
value deduced from the ab initio fit (6.3eV, continuous lines), and a
value enhanced by 20 percent (7.8eV, dotted line). The second computation 
with an  enlarged $U_{3d}$ was made
to obtain an estimate for the 4-band-model shortcomings in comparison
to the fictituous ab-initio result. 

Both correlation functions reduce in strength when electrons are removed.
The longer range 
function does so somewhat stronger. However for the range of interest, i.e.
around optimal doping ($n_d\simeq0.8$), both functions are still 
sizable and not very much smaller than for the
metastable non magnetic half-filled case. This a posteriori justifies
the choice of such a metastable state in the 
ab-initio correlation
calculations. It also demonstrates that for all fillings of interest
very sizable nearest neighbor short range antiferromagnetic correlations
exist together with longer range itinerant antiferromagnetic 
fluctuations.
The $U$-dependence of the longer-range correlation function is more pronounced
and is strongest close to half-filling, indicating again the underlying
Stoner enhancement.

Also of interest is the change of the charge distribution with refilling.
In the single-particle approximation, the electrons close to the Fermi surface
are mostly d-like. Fig. \ref{fig2} displays the non d-fraction of the density 
of states as a function of doping. It can be seen that 
close to the Fermi energy it amounts to 10 percent.
As mentioned before, the model itself is
unable to account for Madelung corrections that would certainly modify
such an extreme density distribution of the removed charge. 
Also a possible redistribution that would come in with the
self consistent computation is not included. However, we will discuss
investigated to which extent
correlations lead to a redistribution of the removed charge.
As can be seen from Fig. \ref{fig2},
there is indeed a sizable change of charcter of the removed density. Most
of this is change into $O 2p$-character. Again, the computation is performed
for two values of $U_{3d}$. The smaller value (6.3eV) leads to a 20 percent
charge redistribution while the larger value leads up to 50 percent
corrections close to half-filling. 
Photoemission experiments not too far from the magnetic state 
found indeed that the electrons removed from the system were largely
of $O 2p$-character \cite{fink}.

\section{Spin correlations and Neutron scattering results}  
                              \label{sec:mod5}

In the ab-initio calculation for the half-filled case,
very strong antiferromagnetic neighbor $Cu$ spin correlations 
in  connection with a $Cu - O$ charge transfer and with longe range  
antiferromagnetic polarizations were found.
 The subsequent model calculations have
shown that the neighbor spin correlations are not restricted to the immediate
vicinity of half-filling but exist for every filling.

This prediction can be tested by comparing the calculated results with
quantitative magnetic neutron measurements. 
From these experiments, a quasi-equal time spin
correlation function $S(\vec{Q})$ was obtained \cite{hayden} for the 
metallic compound $La_{0.85}Sr_{0.15}Cu_2O_4$ 
by extending
the energy integration up to 0.45eV.
The data show a strong longer range structure that is expected
to exist independently of an also found small quasi-elastic
scattering arising from an incommensurate spin density wave
present in this particular compound. 
In the following, a comparison is made between the theoretical equal time 
correlation function and the measured quantity. Both are not
identical. The theoretical quantity is obtained for the 
infinite layer case where for 0.15 holes per unit cell apparently
 no incommensurate spin wave exists,
and consequently less magnetic scattering is expected than for the
measured compound. On the other hand, the theoretical quantity represents 
the true equal time correlation function and contains contributions 
that are not in the range of the measurement.

The limiting equal time case of the measured correlation
function is defined as
\begin{equation}
S(\vec{Q})={1\over N}\int d^3\vec{r}\int d^3\vec{r}'S(\vec{r},\vec{r}')e^{
i\vec{Q}(\vec{r}-\vec{r}')} \ \ \ .
\end{equation}
The theoretical spin correlation function is derived from 
the model calculation at apropriate
doping.
It is represented by spin
correlations between different orthogonal orbitals,             
\begin{equation}
S(i,j,\vec{G})=\langle\Psi_{corr}|\vec{s}_i(0)\vec{s}_j(\vec{G})|\Psi_{corr}\rangle \ \ .
\end{equation}
Here $i$ denotes the i-th orbital in the unit cell with atom position 
$\vec{r}_i$, $\vec{G}$ represents the lattice vectors, and $\vec{s_i}(\vec{G})$
represents the spin operator for orbital $i$ in the $\vec{G}$ unit cell.
When assuming that the spatial moment distribution is shrunk to
the nuclear positions,
\begin{equation}
S(\vec{r},\vec{r}')\simeq \sum_{i,j,\vec{G},\vec{G}'}\delta (\vec{r}-\vec{r}_i+\vec{G})
\delta (\vec{r}'-\vec{r}_j+\vec{G}+\vec{G}')S(i,j,\vec{G}') \ \  ,
\end{equation}
one obtains
\begin{equation}
S(\vec{Q})=\sum_{i,j,\vec{G} } S(i,j,\vec{G})e^{i\vec{Q}(\vec{r}_i-\vec{r}_j-\vec{G})} \ \ \ .
\label{eqs3}
\end{equation}
This function is very easy to compute.
Fig. \ref{figs1} contains the results for a particular $\vec{Q}$ direction, 
namely the diagonal (1,1) axis, obtained in different approximations. 
The zone boundary is at h=1, the intensity is
given per formular unit which here is equivalent to a unit cell or to a
single $Cu$ atom. 

The lowest curve represents the result for the single-particle ground
state. It represents the exchange holes. As the finite value at $h=0$
indicates, the summation in eq. \ref{eqs3} was not brought to convergency.
Instead, the $Cu-Cu$ density matrix elements were only included up to the
4th neighbor, and no density matrix elements with $4s$ or $2p$ orbitals
extending beyound the nearest neighbor $Cu-O$ terms were added. The
maximal deviation occurs for $h=0$ were the contributions from all 
missing terms add up. Due to dephasing, the correction is very much
smaller for finite $h$. 
Due to the fine structure in the unit cell, this function is
finite at the first lattice vector ($h=2$).
This represents the $Cu-O$ correlation function.

Next, short range correlations as they are deduced from a single coherent
5 $Cu$ cluster calculation are included (dotted curve). Here, the nearest
neighbor $Cu-Cu$ correlations come into play and cause a peak at
the zone boundary ($h=1$).
When extending the correlation treatment to a 9 Cu
cluster, second and third neighbor correlations are more correctly treated.
They lead to a narrowing of the peak and to a small enhancement (continuous
curve). Finally, also the corresponding values with enlarged $U$ (7.8eV
instead of 6.3eV) are given (broken curve). Increasing $U$ leads to a
strong enhancement of the maximum, indicating again the proximity to a
magnetic phase transition.
These results are compared to experiment \cite{hayden} (dots) in Fig. 
\ref{figs1}. 
As expected, the
theoretical equal time correlation function is always larger
than the experimental correlation function whose energy integration extends
only to 0.45 eV. 
Beyound h=1.5, the experimental results
are influenced by the next Bragg peak, and no more meaningful. 

There are specific contributions to the theoretical correlation
function that are not expected to be seen by experiment. These
are the short range contributions connected with the $Cu-O$ hopping,
arising already without correlations. The hopping energy connected with
this part 
of the correlation function 
is $t=1.6$eV, and very much larger than the energy cut-off.
Consequently, only a marginal part of these contributions is expected to
show up in experiment. 
A considerably larger fraction of the uncorrelated longer
range $Cu-Cu$ contributions is expected to show up since these mostly
arise from the uppermost band. Also the relevant correlation contributions
are expected to be measured by experiment. While the on-site correlation
functions might not fully show up, the effect of the 
neighbor $Cu-Cu$ spin correlations
is expected to arise mostly from the electrons in the uppermost band,
and the longer range enhanced spin correlations are certainly connected
with electrons close to the Fermi surface, as is indicated by their
strong resonance dependence on $U$. 
Fig. \ref{figs2} displays the correlation functions of Fig. \ref{figs1},
but with all $Cu-O$ contributions of the single-particle approximation
removed, and with the residual function shifted so that 
$S(0)=0$ holds. The residual $Cu-Cu$ single-particle contributions are
small and essentially bell shaped (lowest curve). The correlation 
contributions lead
to a  pronounced maximum around the zone boundary. When correlations
resulting from the 9 $Cu$-cluster calculation are included, 
then the half width of the
correlation peak corresponds well to the half width of the experimental
peak. However, for the value of $U$ taken from the fit to on-site
correlations, the integrated scattering intensity is not larger than
the experimental counterpart. 
The result for a 20 percent enhanced $U$ finally leads to a correlation
function that is systematically larger than the experimental curve.
The ab-initio calculation if performed for the relevant doping
would certainly give a correlation function as large or even somewhat larger
than the model result for the enhanced interaction.  A future comparison
with experimental results for a metallic compound without a spin density wave
will allow to decide whether the model results or the ab-initio results
are more reliable. The ab-initio calculations might overestimate the
Stoner enhancement, while the model might well leave out relevant
degrees of freedom, and might consequently need to be extended.

The theoretical results represent not only the particular
doping of 0.15 holes
but should be representative for a wider range of doping even farther 
away from the
magnetic case. As Fig. \ref{fig10} demonstrates, the neighbor
correlations that represent the weight of the peak around the zone boundary
reduce only slowly with further doping. The longer range correlations
are expected to reduce faster, so that a continuous widening of the
peak with further doping is expected. These correlation features
appear over a rather wide range of doping and have consequently,
no direct connection with any kind of Mott Hubbard transition.

The equal-time spin correlation function was computed earlier for a
one-band Hubbard model \cite{jap1,sca1} or for a t-J model \cite{daga2}.
In these computations, due to
the particular Fermi surface, at 0.15 doping a spin-density wave shows up. The
resulting equal time correlation function is different from the
one given in Fig. \ref{figs1}. It is close to the bell
shaped curve of the uncorrelated electrons in Fig. \ref{figs2} but
enhanced by a factor of 3. In addition, for magnetically ordered states,
there is a very narrow peak just at $h=1.0$ ( or a set of two peaks close
to this point). This peak dissapears for the not ordered states, but
its width is usually not resolved due the finite k-point mesh used
in these computations.
There is no evidence for strong shorter- or longer-range 
spin-correlation features in the nonmagnetic metallic state. This
indicates that single-band models with local interactions don't adequately
describe
the low-energy degrees of freedom of the metallic case.

The extended range of
longer-range antiferromagnetic correlations found in the ab-initio
calculation but also for the 4-band model, contrasts to
single-band model results. This is connected to the following
difference . On-site
correlations are strongest for a $3d_{x^2-y^2}$ occupation of 1.
This occupation occurs at 0.4 to 0.5 doping. Antiferromagnetic
order on the other hand is strongest for perfect nesting in the
half-filled band case. Inbetween both points, a region of strong fluctuations
is found. For a single-band model, these different points are reduced to
a single point, half-filling.

     \section{   Conclusions }  
                              \label{sec:conclu}

The results obtained within the framework of the LA
can be hierachically classified into the following categories. The first
concerns  the general field of ab-initio 
calculations for the transition metals,
the next deals with the connection of full Hamiltonian and model Hamiltonians, 
and the last one finally covers the specific electronic properties of the
high-$T_c$ compounds. 

Concerning the general field of ab-initio calculations for 
transition metals, we have presented 
the first LA computation which also provides detailed
correlation functions, and which, as mentioned is not
connected to a homogeneous-electron-gas like approximation.
We could explicitly investigate  those correlation effects 
that are out of the reach of the LDA.
An example relevant for the general field
is the correlation induced $3d-4s$ charge transfer. A
similar charge transfer is expected from magnetic order in the 
transition metals and was proposed long ago by
Lang and Ehrenreich 
as an explanation
for the inverse magneto-volume effect in $Ni$ \cite{le}. This can now
be quantitatively adressed.

We have found that the
weak-correlation expansion in which the LA is computed
can be successfully applied to systems as strongly correlated
as the high-$T_c$ materials.
Astonishingly, problems arose neither in the context of
strong atomic correlations nor due to a possible Mott-Hubbard transition 
on either the atomic or a more extended unit-cell scale,
but only from  the closeness of the ground state
to a magnetic phase and from the resulting Stoner enhancement. 
A future extension of this weak-correlation
expansion, from the linearized to the full
CCSD equations in the restricted operator space should 
help to improve this specific shortcoming.

The calculations have also demonstrated how important it is
to use an SCF-calculation for the full solid as a starting point.
In the past, it has been necessary to restrict oneself to calculations
for small $Cu-O$ clusters when  applications based on ab-initio treatments
beyound a homogeneous-electron-gas-based
approximation were made.
One of the first such calculations has also been 
done in the framework  of the LA \cite{mest}, and
a comparison of the results
clearly shows how adversely 
 cluster constraints  influence basic results like the electronic 
charge distributions or correlation functions.
\\

Concerning the transition from the ab-initio calculation to a 
model Hamiltonian, the dominant issue is the determination
and the analysis of the interaction parameters of the effective
Hubbard models. Among others, we have provided
a detailed derivation
of the effective $Cu 3d_{x^2-y^2}$-interaction parameters,
starting from the bare Coulomb interaction, and analysed in
particular the screening effect of the $4s,4p$-electrons that had
been proposed long ago by Herring \cite{herr}. 
A surprising finding was that this screening is not much connected with
the residual interactions between the $3d$ and the $4s,4p$ electrons, 
but largely mediated by a charge transfer from the $3d$ orbitals into the
$4s,4p$ orbitals, when starting from the SCF-ground state.

The obtained $Cu 3d_{x^2-y^2}$-interaction parameter turned out to
be somewhat smaller than the global $Cu 3d$-interaction parameters that
were determined by frozen charge LDA calculations.
The difference apparently results from residual
interactions between electrons in $3d_{x^2-y^2}$-orbitals on
neighbor-$Cu$ sites that are only accounted for when
correlation functions are used as a means to determine the interaction 
parameter.
A similar deviation was noted earlier for the case of the
transition-metal interaction parameters, for which somewhat smaller
values were obtained from fits to experiment than 
from computations by the LDA \cite{8}.
\\

Finally, we will  adress
the specific properties of the metallic
$CuO$ compounds. On the single-electron level, our results 
are similar to the LDA results. This concerns in particular 
the relevance of the
$4s$ orbitals for the dispersion of the half-filled band and 
for the form of the Fermi surface. 
However, there are also  differences.
Surprising is the one for the $3d$-occupation which comes out
too large in the LDA.
Furthermore, we found sizable correlations on different length scales.
While the strong
atomic correlations were expected for these compounds, 
we found in addition a strong magnetic nearest-neighbor
$Cu-Cu$ correlation that might even lead to a
neighbor attraction of electrons with different spin. This correlation
is not due the longer-range magnetic fluctuations, however it
may well be enhanced by it.
It is also connected with
a sizable $Cu-O$-charge transfer. A 
homogeneous-electron-gas-based method like the LDA is neither expected
to be able
to handle such a correlation nor the connected charge transfer.
This explains the just-mentioned difference in the $3d$-occupation.
In addition, a sizable long-range magnetic polarization was
found that can best be described in terms of a Stoner enhancement. All 
these
features turn out to be present over a large doping range, and not
only very close to half-filling.

The connection between  variations in the magnetic correlations
and the  charge transfer is expected to  result
in interesting couplings between the magnetic and lattice degrees of
freedom. 
In particular, it should not be surprising if the magneto-volume effect
for the case of the
half-filled-band systems turns out to be very small or
even negative, as is the case in $Ni$.

The just-mentioned particular neighbor
spin correlations, enhanced by
longer range magnetic fluctuations, dominate the spin correlation
function, and explain the features found in the
measured spin correlations for
$La_{0.85} Sr_{0.15} Cu_2 O_4$ \cite{hayden}.
Apparently, neither the 
single-band-Hubbard model results nor the t-J-model results can explain 
this spin correlation function (see discussion in section \ref{sec:mod5}).
It was  not even possible to bring the results of the 4-band model 
with only on-site interactions  
to good agreeement with the ab-initio results, when
on-site and longer-range correlations were jointly concerned.
It seems that a proper description of the real system can only be obtained
if in such a 4-band model the background-induced
magnetic $Cu$ neighbor interactions
are explicitly taken into account, or if the  model is 
generalized by an  explicit
inclusion of the $4p$ orbitals, may be even of the full 
screening process of the $3d$-interactions
by the $4s,4p$-orbitals. In our future work, we shall investigate
such extensions.

To conclude, ab-initio correlation calculations can now be
performed for the
transition metals.
With the local ansatz, details of the correlation functions as well as a  
good understanding of the relevant short-range correlation features can be 
obtained. The first application for a metallic  high-$T_c$ compound
shows a fairly good agreement between the computed and the
measured magnetic correlation functions.

\section*{Acknowledgements}
Thanks are due to C. J. Mei for making available
the input files of his earlier calculations, and in particular
to G. Zwicknagl, B. Farid, and O. Gunnarrson
 for a careful reading of the manuscript.

\onecolumn
\begin{table}
{
\begin{center}
{   \begin{tabular}{|l|rrr|}
 \hline                                                                        
Correlations&$E_{corr}$(a.u./u.c.)&spin contrib.&\quad\\
\hline
on site & -.1248 &&\\
Cu-O nn & -.0154& 0.20&\\
Cu-Cu nn & -.0190& 0.80&\\
O-O nn& -.0033&&\\
Cu-O nnn& -.0036&&\\
Cu-Cu nnn & -.0157& 0.95&\\
Cu-Cu nnnn& -.0040& 0.95&\\
\hline
\end{tabular}}                                                                 
\end{center}}\protect
\caption{\protect
Correlation energy contributions in atomic units
for particular succesively added operators.}
\label{tab31}
\end{table}                                                        

\begin{table}
\begin{center}
{   \begin{tabular}{|l|r|rrr|lc|}
 \hline                                                                        
Orbital& HF&on site corr&nn corr&full corr&LDA&\quad\\
\hline
$Cu 3d_{x^2-y^2}$ & 1.51& 1.33& 1.17& 1.15&&\\
$Cu 3d_{z^2}$ & 1.95& 1.94& 1.94& 1.94& 9.30$^*$&\\
$Cu 3d_{xy},3d_{xz},3d_{yz}$ & 2.00& 2.00& 2.00& 2.00&&\\
$Cu 4s$ & 0.50& 0.55& 0.57& 0.58& 0.53 &\\
$Cu 4p_{pl}$ & 0.30& 0.33& 0.34& 0.34&&\\
$Cu 4p_{\perp}$ & 0.09& 0.11& 0.11& 0.11 
&\raisebox{1.5ex}[-1.5ex]{0.64$^*$}&\\
$O 2s$ & 1.82& 1.82& 1.81& 1.81&&\\
$O 2p_b$ & 1.42& 1.48& 1.57& 1.58&&\\
$O 2p_{orth}$ & 1.97& 1.96& 1.96& 1.96&&\\
$O 2p_{\perp}$ & 1.95& 1.93& 1.92& 1.91&&\\
\hline
\end{tabular}}                                                                 
\end{center}\protect
\caption{
Charge distributions for the SCF ground state and with correlations
added, in comparison to LDA results \protect\cite{kgj}.
$\ ^*$ The sums over five respectively three partial contributions 
each are given.}
\label{tab32}
\end{table}

\begin{table}
{
\begin{center}
{   \begin{tabular}{|l|r|rrrr|}
 \hline                                                                        
Orbital& $n_i$& $\Delta(HF)$&$\Delta(corr)$&$\Delta_{max}(corr)$&\quad\\
\hline
$Cu 3d_{x^2-y^2}$ & 1.17& 0.243& -0.122& -0.172&\\
$Cu 4s$ & 0.57& 0.203& -0.009& -0.081&\\
$Cu 4p_{pl}$ & 0.34&0.141& -0.005& -0.029&\\
$O 2s$ & 1.81& 0.086& -0.002& -0.009&\\
$O 2p_b$ & 1.57& 0.168& -0.033 &-0.049&\\
\hline
\end{tabular}}                                                                 
\end{center}}\protect
\caption{
On-site correlations for the different atomic orbitals. The individual 
terms are defined in the text.}
\label{tab33}
\end{table}

\begin{table}
{
\begin{center}
{   \begin{tabular}{|l|rrrr|}
 \hline                                                                        
Included correlations&$\nu=0$&$\nu=1$&$\nu=2$&$\nu=3$\\
\hline
HF ground state&0.276&-0.012&0.001&0.001\\
\hline
on site& 0.478&&&\\
up to $\nu=1$& 0.530& -0.140&&\\
up to $\nu=2$& 0.540& -0.220& 0.170&\\
up to $\nu=3$& 0.543& -0.243& 0.183& 0.069\\
\hline
Heisenberg model& 0.75\ & -0.34\ & $>$0.10\ & $>$0.10\ \\
\end{tabular}}                                                                 
\end{center}}\protect
\caption{\protect
Spin correlation functions for $Cu 3d_{x^2-y^2}$-orbitals between
neighbor sites $i,i+\nu$,
as functions of the included correlation operators, in comparison 
to the 2d-Heisenberg model.}
\label{tabss}
\end{table}

\begin{table}
{
\begin{center}
{   \begin{tabular}{|l|rrrr|rr|}
 \hline                                                                        
Correlations& $U_d$&&$U_s$&$U_{sd}$&$U_p$&\\
\hline
on site, single without charge transfer & 20.8&& 1.8& 1.3
& 6.6&\\
on site,single & 12.0&& 1.7& 1.1& 6.6&\\
on site, global& 10.4&& 1.8& 1.2& 8.0&\\
$CuO$-nn, global& 8.8&& 1.8& 1.1& 7.6&\\
all, global & 6.3&& 1.8& 1.1&&\\
\hline
estimate& 5.7& $\pm$1.0& 1.8& 1.1& 6.0& $\pm$1.0\\
LDA& 9.0& $\pm$1.0&&& 4.5& $\pm$2.0\\
\end{tabular}}                                                                 
\end{center}}\protect
\caption{
Effective on-site interaction parameters $U_i$ (eV), as
obtained in different approximations. The final estimate is also
given in comparison to typical LDA results \protect\cite{mms,sj,hyb,mam}.
The individual terms are explained in the text.}
\label{tabmod1}
\end{table}                                                        

\begin{table}
{
\begin{center}
{\begin{tabular}{|l|r|rrr|}
\hline                                                                        
Computation& $\Delta_{3d}$&&$\vec{S_i}\vec{S_{i+\nu}}$&\\
&&$\nu=1$&$\nu=2$&$\nu=3$\\
\hline
ab initio & -0.122& -0.140& 0.075& 0.072\\
\hline
$U_{d}=6.3$eV& -0.122& -0.071& 0.013& 0.011\\
$U_{d}=7.8$eV& -0.145& -0.097& 0.025& 0.024\\
$U_{d}=9.4$eV& -0.160& -0.126& 0.045& 0.043\\
\end{tabular}}                                                                 
\end{center}}\protect
\caption{On site and $\nu$th neighbor
$Cu-3d_{x^2-y^2}$ correlation functions
 for the cluster with 5 active $Cu$-atoms.
Ab initio results in comparison to model results with differing $U_{d}$} 
\label{tabmod2}
\end{table}

 
\begin{figure}
\protect\caption{\protect Schematic representation of the two largest
clusters
for which interaction matrix elements for the basis orbitals were
  generated. $Cu$-atoms are denoted by crosses, $O$-atoms for which
correlation operators were included are denoted by filled circles,
while $O$ atoms who contributed only to the $V_{ijkl}$ are denoted by
open circles.}
\label{fig1} 
\end{figure}

\begin{figure}
\protect
\caption{\protect Nearest neighbor ($\nu=1$) and second nearest neighbor ($\nu=2$)
Spin correlation function in dependence of the filling $n_b$ of
the uppermost band, obtained for a model with $U_{3d}$=6.3 eV
(continuous lines) or 7.8 eV (broken curve), respectively.}
\label{fig10}
\end{figure}

\begin{figure}
\protect
\caption{\protect Relative non 3d-like density of states at the Fermi energy 
 in dependence of the filling $n_b$ of
the uppermost band, obtained without correlations (dotted-broken curve),
for  $U_{3d}$=6.3 eV
(continuous lines) and 7.8 eV (broken curve), respectively.}  
\label{fig2}
\end{figure}

\begin{figure}
\protect
\caption{\protect Equal time spin correlation function 
S(Q) for $\vec{Q}=(h,h,0)$
in comparison to experiment \protect\cite{hayden} (empty circles).
Given are the results of the HF-ground state (broken-dotted curve),
the 5 atom cluster result (dotted line) and the 9 atom cluster result
(continuous line) for  $U_{3d}$=6.3eV, and the 9 atom 
cluster result for
$U_{3d}$=7.8eV.}
\label{figs1}
\end{figure}

\begin{figure}
\protect
\caption{$Cu-Cu$ -dependent part of the equal time spin correlation
function S(Q) in comparison to experiment. Definition of the
curves like in fig.~\protect\ref{figs1}}
\label{figs2}
\end{figure}

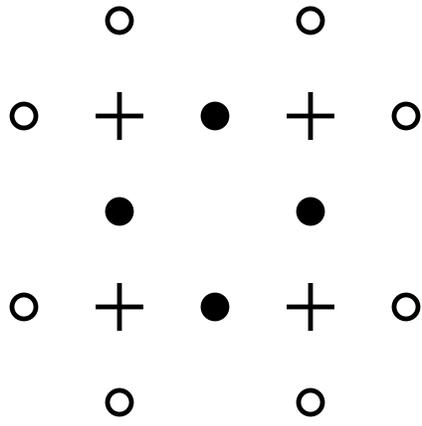
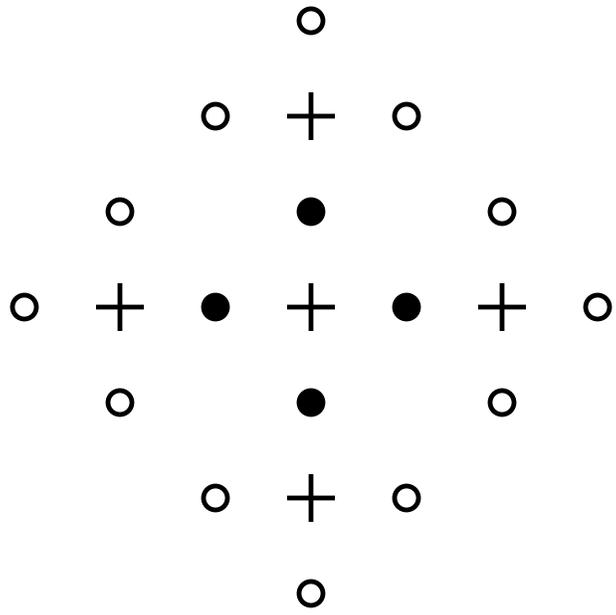
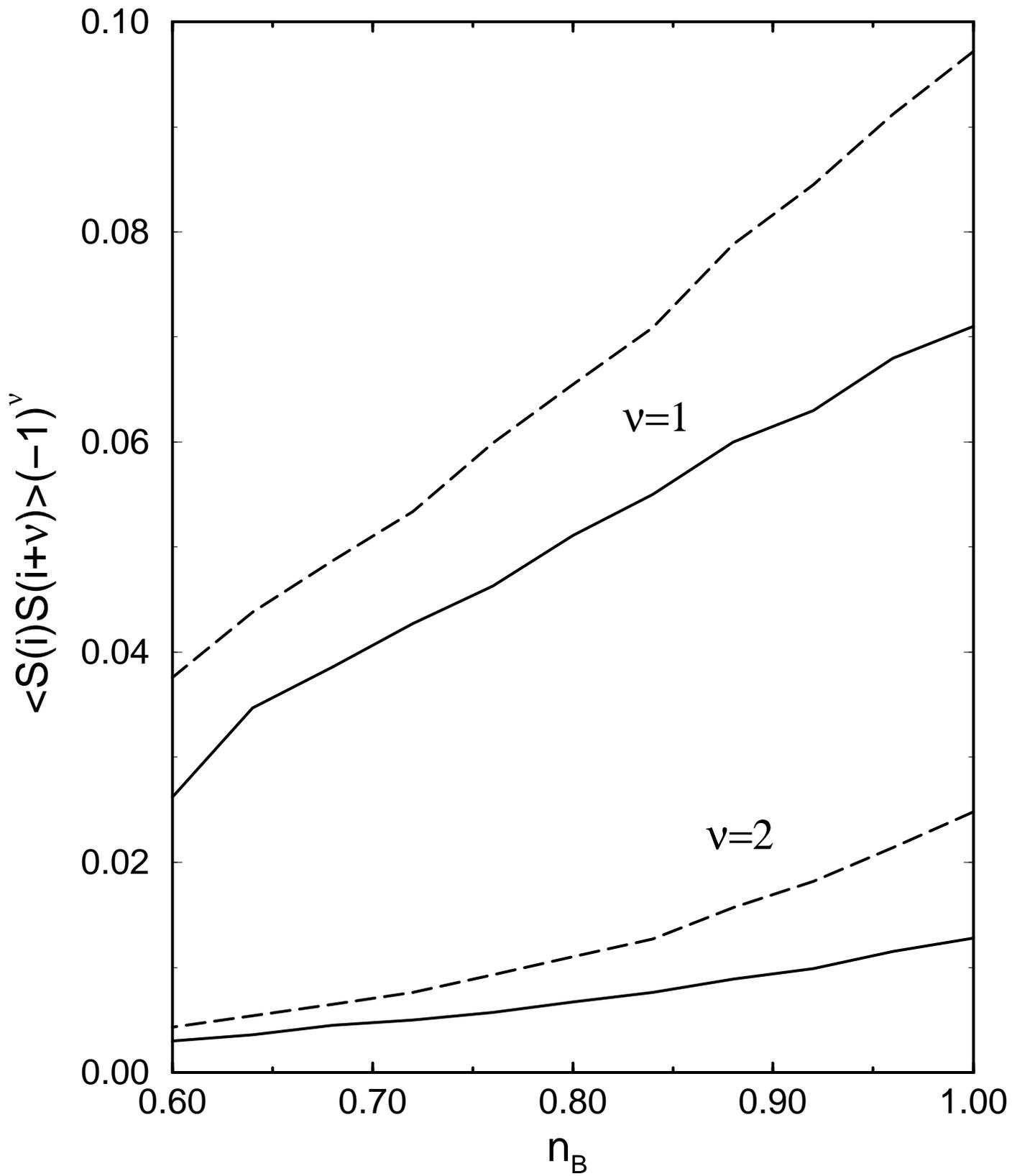
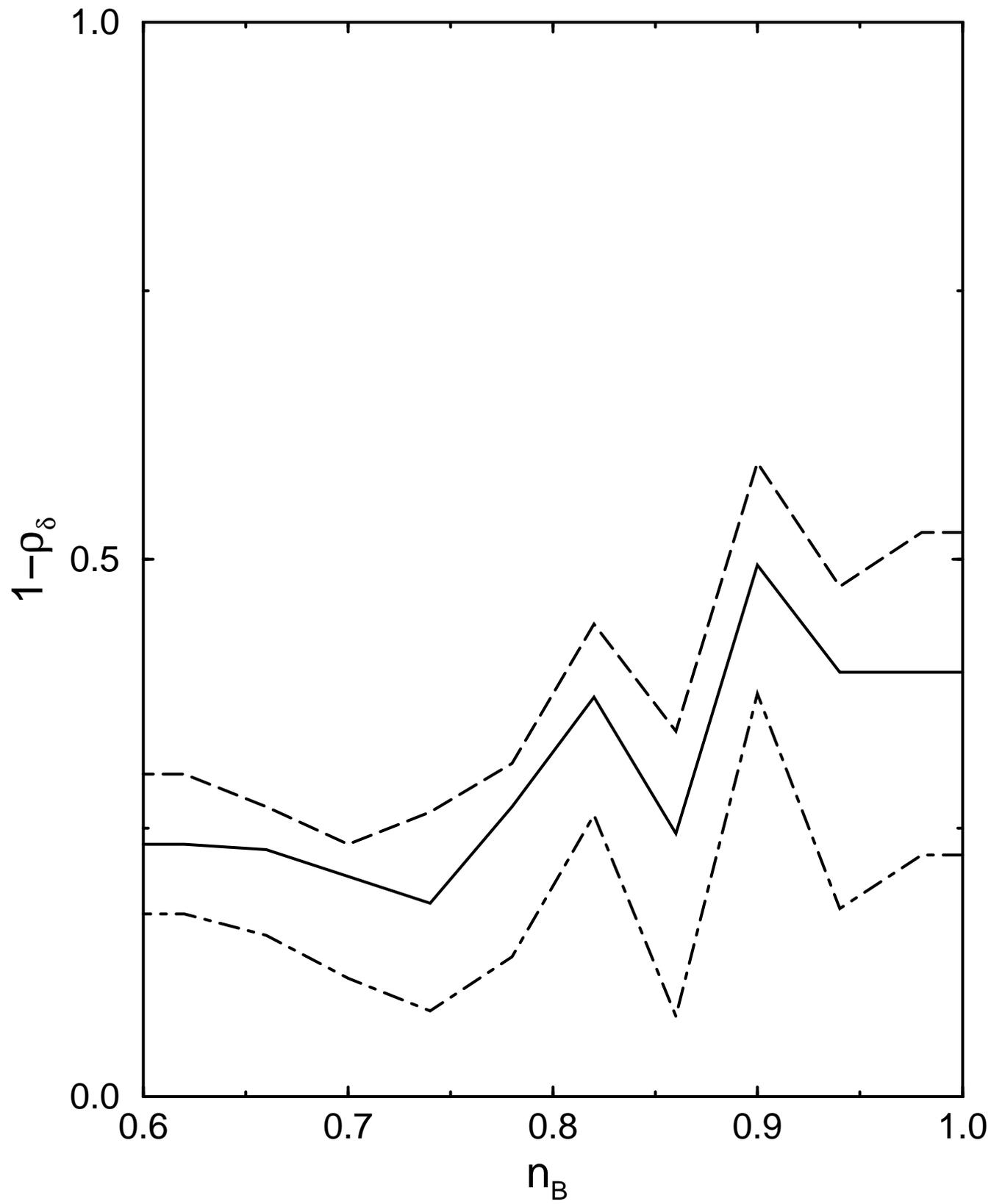
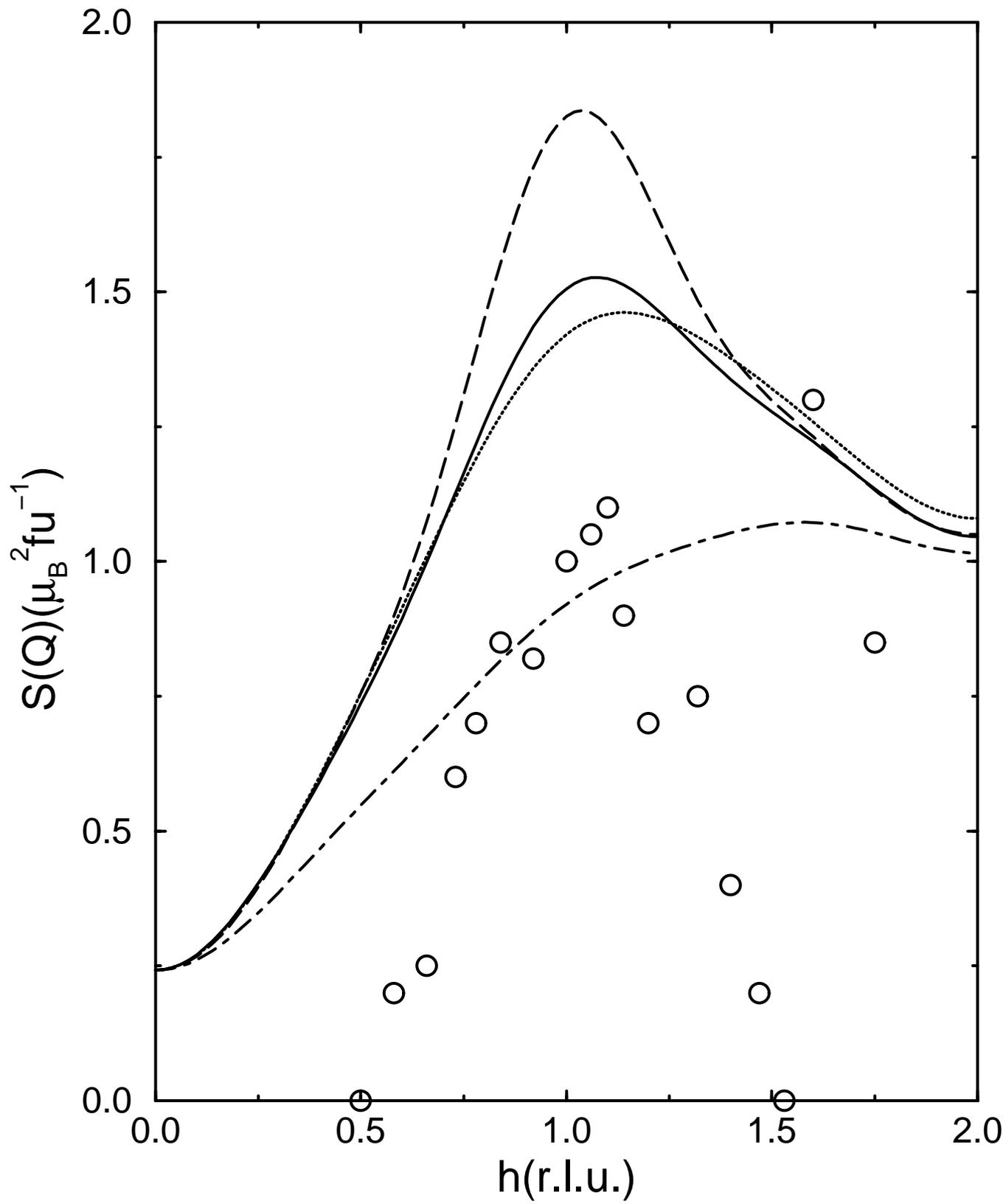
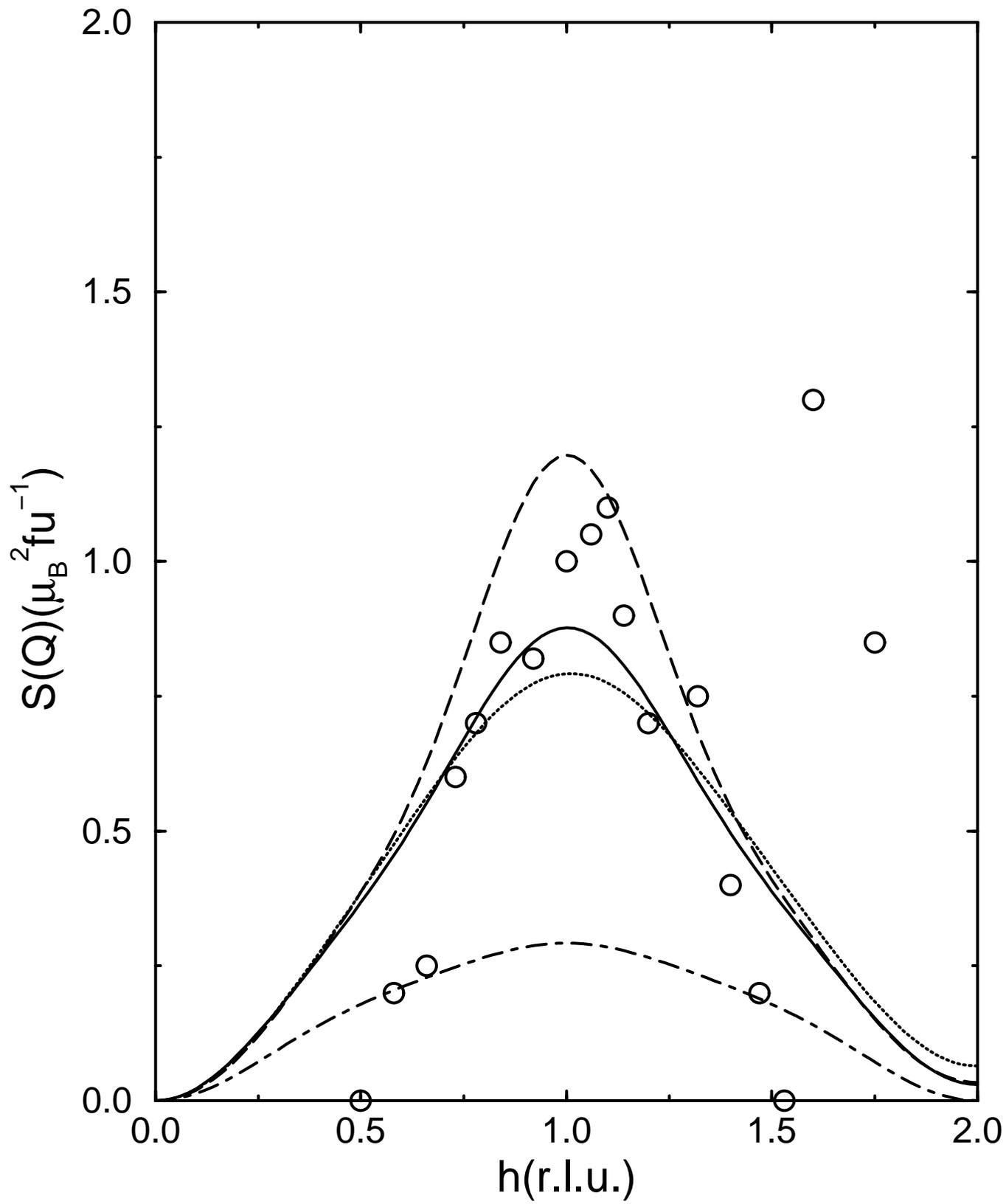
\end{document}